\documentclass[aps,prx,showpacs,twocolumn,floats,epsfig,longbibliography]{revtex4-2}
\usepackage{amssymb}
\usepackage{amsbsy}
\usepackage{amsmath}
\usepackage{xcolor}
\usepackage{enumerate}
\usepackage{bm}
\usepackage{epsfig}
\usepackage{makecell}
\usepackage{color}
\usepackage{soul}
\usepackage{changepage}
\usepackage{float}
\usepackage{braket}
\usepackage[colorlinks,linktocpage,bookmarks=false,citecolor=blue,linkcolor=red,urlcolor=blue]{hyperref}
\newcommand\mybar{\kern1pt\rule[-\dp\strutbox]{.8pt}{\baselineskip}\kern1pt}

\newcommand{\beq}{\begin{equation}}
\newcommand{\eeq}{\end{equation}}
\newcommand{\bea}{\begin{eqnarray}}
\newcommand{\eea}{\end{eqnarray}}

\newcommand{\Tr}{{\rm Tr}}

\usepackage{xcolor}

\begin{document}
\title{Separability criterion using one observable for special states: Entanglement detection via quantum quench}
\author{Roopayan Ghosh}
\affiliation{Department of Physics and Astronomy, University College London,Gower Street, WC1E6BT, London}
\author{Sougato Bose}
\affiliation{Department of Physics and Astronomy, University College London,Gower Street, WC1E6BT, London}
\begin{abstract}
Detecting entanglement in many-body quantum systems is crucial but challenging, typically requiring multiple measurements. Here, we establish the class of states where measuring connected correlations in just {\em one} basis gives a necessary condition to detect bipartite separability  thus offering a sufficient condition to detect entanglement, provided the appropriate basis and observables are chosen. This methodology leverages prior information about the state, which, although insufficient to reveal the complete state or its entanglement, enables our one basis approach to be effective. We discuss the possibility of one observable entanglement detection in a variety of systems, including those without conserved charges, such as the Transverse Ising model, reaching the appropriate basis via quantum quench. This provides a much simpler pathway of detection than previous works. It also shows improved sensitivity from Pearson Correlation detection techniques.
\end{abstract}
\maketitle
\paragraph*{\textbf{Introduction:}}
Detection of many-body entanglement (bi-partite or multipartite) is an experimentally difficult task insofar only a handful of experiments \cite{Islam2015,zollerscience.aau4963} have been able to successfully achieve it. The typical methods involve creation of a replica of the original many-body system to compute purity\cite{Islam2015} or using a series of randomized measurements to detect Renyi entropy in a trapped ion quantum simulator\cite{zollerscience.aau4963}. Both of the set-ups were fine tuned and while they can measure the entanglement of a carefully prepared state in an idealized setting, measuring  or witnessing entanglement of states obtained in many-body experiments simply and effectively is practically impossible\cite{PhysRevLett.129.230503}.

 That being said, entanglement detection in general many-body quantum systems is crucial for several reasons. Firstly, numerous emergent many-body systems in solid-state structures, such as multiple electrons confined to nanowires, can be described by at least one of the Hamiltonians discussed in this work. Additionally, various other types of Hamiltonians can be realized in engineered quantum dot arrays. Detecting entanglement between different parts of such systems is vital for certifying the genuine quantum nature of these systems. Moreover, entanglement between two components of a many-body system has been a subject of study for an extended period, with quantifications such as von Neumann entropy and negativity being employed. The motivation behind these studies was to comprehend the entanglement structure of such systems, either to facilitate effective variational descriptions or to assess the natural availability of resources for quantum communication. However, one needs to actually verify this entanglement in experiments, and the above quantifications are difficult to implement experimentally. 

Hence it is of paramount importance to connect entanglement detection in specific classes many-body states with quantities requiring less demanding experimental resources to measure. To begin, let us look at the most general definition of an entangled state.
An entangled state is defined as one that cannot be separated into individual states belonging in separate subspaces. In the simplest scenario of bipartitions, i.e., a system with constituent Hilbert spaces $\mathcal{H}_A$ and $\mathcal{H}_B$, and the combined space given by $\mathcal{H} = \mathcal{H}_A \otimes \mathcal{H}_B$, a pure state $|\psi\rangle \in \mathcal{H}$ is separable iff,
\begin{equation}
    |\psi\rangle=|\psi_A\rangle \otimes |\psi_B\rangle
    \label{eq:pureseparable}
\end{equation}
where $|\psi_{A[B]}\rangle \in \mathcal{H}_{A[B]}$. On the other hand, a mixed state represented by density matrix $\rho$ is separable iff
\begin{equation}
    \rho=\sum_ip_i\rho_i^A \otimes \rho_i^B
     \label{eq:mixedseparable}
\end{equation}
where $\rho_i^{A[B]} \in \mathcal{H}_{A[B]}$ and $\sum_i p_i=1$.

Entanglement indicates quantum correlations, and measurement of classical correlations between one set of observables between the corresponding subsystems is typically not enough to detect entanglement --- classical correlations should necessarily exist in  several different observables. The search for detection of entanglement with least number of measurements have taken us in various directions in recent years\cite{PhysRevA.70.062113, PRXQuantum.3.010342,DeChiara_2018,PhysRevLett.129.260501,Neven2021,10.21468/SciPostPhys.12.3.106,PhysRevLett.125.200502,gray2018machine,banchi2016entanglement,squillante2023gr}, with specific techniques available for spin systems\cite{Cirac1,PhysRevB.75.054422,PhysRevResearch.5.013158,PhysRevA.71.010301,PhysRevA.72.032309,PhysRevB.82.012405}, measuring charge fluctuations \cite{Petrescu_2014,PhysRevB.85.035409,PhysRevB.83.161408,PhysRevB.107.014308}, coherences \cite{PhysRevA.99.052354} or number entropy\cite{Lukinscience.aau0818,PhysRevLett.124.243601,PhysRevB.105.144203,PhysRevLett.130.136201} in systems with locally conserved charge, and measurements of classical correlations in a set of Mutually Unbiased Bases(MUBs) or observables\cite{PhysRevA.101.022112,PhysRevA.86.022311,Maccone1,Macconemulti,Erker2017quantifyinghigh,Hiesmayr_2021,sadana2022relating}.


The focus of this work shall be on the efficient detection of bipartite entanglement utilizing the aforesaid MUBs. Previous studies involving MUBs have revealed that at least two measurements are necessary to establish a reliable criterion for detection. However, considering that bipartite entanglement typically implies classical correlation in  many measurements,  naively we expect the existence of at least one measurement that exhibits zero correlation for separable states, but non-zero correlation for entangled states. Then by measuring this specific quantity, our problem of entanglement detection should be resolved. But the challenge lies in the fact that there is no guarantee of existence of such a quantity that is independent of the state being examined, thus leading to a wide variety of state-dependent entanglement witnesses. 

In this work we find a common observable which can efficiently detect bipartite entanglement for a wide class of states. We demonstrate that, (i) it is in fact possible to detect entanglement of any arbitrary mixed state via measuring correlations of one observable independently of the measured state, if they are separable to purely real states or  states with the density matrix having purely imaginary off diagonal elements. (ii) We also show that in presence of a locally conserved charge such as particle number/magnetization, detection of entanglement in any generic state (i.e. need not be separable to purely real or imaginary) is possible by this technique. \footnote{There are several methods already available for entanglement detection in this case,but mostly focusing on pure states. \cite{Petrescu_2014,PhysRevB.85.035409,PhysRevB.83.161408,PhysRevB.107.014308,PhysRevA.99.052354,Lukinscience.aau0818,PhysRevLett.124.243601,PhysRevB.105.144203,PhysRevLett.130.136201}. However this method provides a new correlation witness from the point of view of MUBs for all kinds of states.}
 (iii) We then test the efficacy of the observable to detect entanglement in mixed states generated by different Hamiltonians via exact numerical simulations. 
 
In what follows, since our proposed criterion is based on MUB, we first define MUBs and discuss a specific entanglement witness which utilizes them. Then we elaborate our criterion and provide examples comparing efficiency of both approaches.

\paragraph*{\textbf{Mutually Unbiased Basis(MUB):}}
A set of bases $\{\bm{\alpha}\}=\{\ket{\alpha_i}\}$ and $\{\bm{\alpha^{\prime}}\}=\{\ket{\alpha^{\prime}_j}\}$ where $i,j \in 1,\hdots ,d$, in a Hilbert space $\mathcal{H}=\mathbb{C}^d$ is mutually unbiased iff,
\begin{equation}
|\langle\alpha_i|\alpha^{\prime}_j\rangle|^2=1/d \hspace{0.5 in}   \forall \hspace{0.2 in}  i,j
\label{MUB construction}
\end{equation}

\paragraph*{\textbf{Entanglement witnessing using MUBs}}
Entanglement detection using MUBs requires measurement of correlations. Among the several different quantifiers which have been addressed in previous works \cite{PhysRevA.101.022112,PhysRevA.86.022311,Maccone1,Macconemulti,Erker2017quantifyinghigh,Hiesmayr_2021}, we focus on the Pearson correlation measurement, as it was shown to be among the more efficient measures.\cite{Maccone1} 

Consider a system $S$, subdivided into two subsystems $A$ and $B$. Let us denote an observable measured in subsystem $A$ as $\mathcal{O}_A$ and for $B$ as $\mathcal{O}_B$. Then the Pearson correlation between the observables is given by the normalized connected correlation,
\begin{equation}
\mathcal{P}_{\mathcal{O}}=\frac{\langle\mathcal{O}_A \otimes \mathcal{O}_B\rangle-\langle\mathcal{O}_A\otimes \mathbb{I}\rangle\langle \mathbb{I}\otimes\mathcal{O}_B\rangle}{\sqrt{\langle\mathcal{O}_A^2\otimes \mathbb{I}\rangle-\langle\mathcal{O}_A\otimes \mathbb{I}\rangle^2}\sqrt{\langle\mathbb{I}\otimes \mathcal{O}_B^2\rangle-\langle \mathbb{I}\otimes\mathcal{O}_B\rangle^2}},
\end{equation}
where the expectation value is taken with respect to the state of the system $S$. For MUB measurement, if we assume that the set of eigenvectors for $\mathcal{O}_A$ is $\{\ket{\alpha}\}$, then we can find another observable $\mathcal{O}^{\prime}_A$ with eigenvectors $\{\ket{\alpha^{\prime}}\}$ fulfilling Eq.~\ref{MUB construction}. A similar construction can be done for subsystem $B$. Then
\begin{equation}
|\mathcal{P}_{\mathcal{O}}|+|\mathcal{P}_{\mathcal{O}^{\prime}}|>1
\label{eq:Macconecond}
\end{equation}
provides a sufficient but not necessary condition for the two halves of the system to be entangled~\cite{Maccone1}. Further improvements on the bound were seen if correlations were measured in more than two MUBs. 

Hereafter, we shall focus in the opposite direction, namely, we shall establish the case where measurements in two basis are not needed, i.e., non-zero connected correlations in one basis immediately indicates entanglement.
 \paragraph*{\textbf{One measurement entanglement detection:}}
 For completeness, we shall discuss the cases of both pure and mixed states,
 \begin{enumerate}[(I)]
     \item  \textbf{Pure states}: In the study of many-body physics we often work with pure states, when we isolate the experimental set-up from interactions with the environment. In this case,
     \begin{equation}
         \mathcal{C}_1=\langle\mathcal{O}_A \otimes \mathcal{O}_B\rangle-\langle\mathcal{O}_A\otimes \mathbb{I}\rangle\langle \mathbb{I}\otimes\mathcal{O}_B\rangle
         \label{eq:C1}
     \end{equation}$=0$, is a \textbf{necessary but not sufficient} condition for zero entanglement between the partitions. This indicates a non zero value of $C_1$ is a sufficient condition for entanglement.
     
     The proof of necessity directly follows from Eq.~\eqref{eq:pureseparable}. The lack of sufficiency condition for separability can be seen from special choices of pure states. For example, if we take the entangled  singlet state $\frac{1}{\sqrt{2}}(\ket{\uparrow\downarrow}-\ket{\downarrow \uparrow})$ and take $\mathcal{O}_A=\sigma_z$ and $\mathcal{O}_B=\sigma_x$, we shall get no correlation.

     \item \textbf{Mixed states}: For generic mixed states, $\mathcal{C}_1=0$ is \textbf{neither a necessary nor a sufficient} for separability. To illustrate this, consider a separable state $\rho_{CC}$ given by,
     \begin{equation}
      \rho_{CC}=\frac{1}{d}\sum_{i=1}^d \ket {\alpha_i}\bra{\alpha_i}\otimes \ket{\beta_i}\bra{\beta_i}
     \end{equation}
where $\{|\beta\rangle\}$ is the eigenbasis of $\mathcal{O}_B$. In this case, $\mathcal{C}_1$ can show high correlations, but
\begin{equation}
    \mathcal{C}_2=\langle\mathcal{O}^{\prime}_A \otimes \mathcal{O}^{\prime}_B\rangle-\langle\mathcal{O}^{\prime}_A\otimes \mathbb{I}\rangle\langle \mathbb{I}\otimes\mathcal{O}^{\prime}_B\rangle
         \label{eq:C2}
\end{equation} 
 will be equal to $0$.

For the generic separable mixed state of Eq.~\eqref{eq:mixedseparable} correlations may exist in multiple bases, but there is no entanglement present. This can be seen by choosing appropriate $\mathcal{O}_A$ and $\mathcal{O}_B$ to compute,
\begin{align}
\langle\mathcal{O}_A \otimes \mathcal{O}_B\rangle &= \sum_i p_i \Tr[\rho_i^A \mathcal{O}_A] \Tr[\rho_i^b \mathcal{O}_B] \nonumber \\
\langle\mathcal{O}_A\otimes \mathbb{I}\rangle\langle \mathbb{I}\otimes\mathcal{O}_B\rangle&=
\sum_i p_i \Tr[\rho_i^A \mathcal{O}_A] \sum_j p_j \Tr[\rho_j^b \mathcal{O}_B]
\label{eqn:mixedcorr}
\end{align}
If $p_i=1$, i.e., the system is separable to $\rho=\rho^A \otimes \rho^B $, then $\mathcal{C}_1=\mathcal{C}_2=0$, and the condition becomes \textbf{necessary}. For all other typical cases it is not. Thus, entanglement witness bounds usually involve computation of correlations in several measurements.
 \end{enumerate}

To overcome this difficulty, we attempt to construct a common observable for all density matrices whose measurement shall provide the \textbf{necessary} criterion. As mentioned before, a common observable is not guaranteed to exist for arbitrary density matrices, so we search for classes of states which allow us to do so.


 
\paragraph*{\textbf{Construction of the observable:}} Let us consider the case of L-qubits, i.e. Hilbert spaces of dimension $2^L$. The generalization to qudits is provided in Appendix~\ref{app:appB}. From the previous discussion of $\rho_{CC}$, choosing an observable whose eigenbasis is mutually unbiased to $\{|\alpha\rangle,|\beta\rangle\}$ is clearly necessary (For simplicity we hereafter take $\{|\alpha\rangle,|\beta\rangle\}$  as the computational $\sigma_z$ basis of the respective systems). Additionally, for separable states not of the form of $\rho_{CC}$, any arbitrary mutually unbiased observable does not yield $\mathcal{C}_2=0$.  Ideally since we also want an observable easily implementable in experiments, the special observable should to be realized by very local (involving on-site or nearest neighbour) transformations. While this is not easily realizable for a collection of qudits, in the L-qubit scenario, a special observable whose eigenbasis is mutually unbiased to the computational basis can be created by rotating $\mathcal{O}_{A[B]}$ via
\begin{equation}
    U_{A[B]}=(e^{-i \sigma^x \pi/4})^{\otimes L_{A[B]}}
    \label{eq:basis1},
\end{equation}
 where $L_{A[B]}$ is the length of subsystem $A[B], L_A+L_B=L$. 

By explicit construction , we see $U$ has the following properties.

\begin{enumerate}[(i)]
    \item $(U_{A[B]})_{p,q}=\frac{1}{\sqrt{d_{A[B]}}}e^{-i \pi(\phi_{p,q})/2}$, {\rm where} $\phi_{p,q} \in [0,L_{A[B]}] \cap \mathbb{Z}$\\
    \item $U_{A[B]}^{T}=U_{A[B]} \implies \phi_{p,q}=\phi_{q,p}$.\\
    \item $(U_{A[B]})_{p,q} (U_{A[B]})_{d-p+1,q}=\frac{1}{d_{A[B]}}e^{-i \pi L_{A[B]}/2} \implies \phi_{p,q}+\phi_{d-p+1,q}=L_{A[B]}$     
\end{enumerate}

 where $d_{A[B]}=2^{L_{A[B]}}$, the dimensions of the subsystem. These properties can be extracted from the tensor product structure of $U_A$ for any $L_A$, by building from the $L_A=1$ case, which in the $\{\ket{0}, \ket{1}\}$ basis is,
\begin{equation}
 U_A=   \left(
\begin{array}{cc}
 \frac{1}{\sqrt{2}} & -\frac{i}{\sqrt{2}} \\
 -\frac{i}{\sqrt{2}} & \frac{1}{\sqrt{2}} \\
\end{array}
\right)
\label{eq:onesite}
\end{equation}
A detailed explanation is provided in Appendix.~\ref{app:appA}

Moving to the Schroedinger picture we obtain the density matrix after the evolution,
\begin{widetext}
 \begin{equation}
 \rho^{\prime}_{i_A, i_B, j_A, j_B}= (U_A \otimes U_B)_{i_A, i_B, k_A, k_B} \rho_{k_A, k_B, l_A, l_B} (U_A \otimes U_B)^{\dagger}_{l_A, l_B, j_A, j_B}
 \label{eq:rhoprime}
 \end{equation}
\end{widetext}

We see that the connected correlator with respect to $\rho^{\prime}$ is identically $0$, when the following conditions are fulfilled,\\ 

\textbf{Condition (i)}
\begin{enumerate}
\item $\rho$ is separable (i.e. given by Eq.~\eqref{eq:mixedseparable}) to purely real [ or with purely imaginary off-diagonal elements] density matrices, i.e. $\rho_i^{A[B]} \in \mathbb{R}^{d_{A[B]} \times d_{A[B]}}$  or $(\rho^{A[B]}_i)_{k,l,k \neq l} \in i\mathbb{R}^{d_{A[B]} \times d_{A[B]}} $.
\item The observable $\mathcal{O}$ is diagonal in the computational basis\footnote{This means that if we want to measure bipartite entanglement not in real space but any other space, for example in momentum space, we have to choose the observable to be diagonal in the computational basis of that space.} with eigenvalues $E_j=\{f(j)+c_1 \ni f(j)=-[+]f(d-j+1), j\in[1,d]\cap \mathbb{Z}, c_1 \in \mathbb{R}\}$. Examples of such observables include total particle number or total magnetization in many-body systems, or the $S_z$ index in large spin systems, all typically measurable by local measurements. The `$[+]$' is required for imaginary case of the previous condition. 
\end{enumerate}
\begin{center}
   \textbf {OR}
\end{center}

\textbf{Condition (ii)}\\
\begin{adjustwidth}{2.5em}{0pt}
$\rho$ is separable with a locally conserved charge, such as magnetization or particle number, and we perform subsystem measurements of the conserved charge. 
\end{adjustwidth}

\paragraph*{\textbf{Proof of condition (i):}}We first show our conjecture is valid when  the first two statements are fulfilled. Dropping the indices $i$ from Eq.~\eqref{eq:mixedseparable} and subscript $A$ from $d_A$ for brevity, we see that,
\begin{align}
\begin{split}
&\Tr[\rho^{\prime A} \mathcal{O}_A] = \sum_{p=1}^{d/2}\sum_{q,r}^d (f (p) +c_1) (U_A)_{p,q} \rho_{q r} (U_A)^*_{r,p} \\
&+ \sum_{p=1}^{d/2}\sum_{q,r}^d (-f (p) +c_1) (U_A)_{d-p+1,q} \rho_{q, r} (U_A)^*_{r,d-p+1}
\end{split}
\label{eq:intermediatesigma}
\end{align}
 
Then using the properties of $U_A$  we can simplify Eq.~\ref{eq:intermediatesigma} by realizing that,
\begin{eqnarray}
&&\sum_{p=1}^{d/2}\sum_{ q,r}^d  f(p) (U_A)_{d-p+1,q} \rho_{q,r}(U_A^*)_{r,d-p+1}\nonumber \\
 &&=\frac{1}{d}\sum_{p=1}^{d/2}\sum_{ q, r}^d  f(p) e^{-i \pi/2[(L_A-\phi_{p,q})-(L_A-\phi_{r,p})]} \rho_{q,r} \nonumber \\
&&=\frac{1}{d}\sum_{p=1}^{d/2}\sum_{ q, r}^d  f(p) e^{-i \pi/2(\phi_{p,q}-\phi_{r,p})} \rho_{q,r} \nonumber \\
&&=\sum_{p=1}^{d/2}\sum_{ q,r}^d  f(p) (U_A)_{p,q} \rho_{q,r}(U_A^*)_{r,p} 
\label{eq:proof}
\end{eqnarray}
where in the third step, we have used $U_A^T=U_A$ and $\rho_{q,r}=\rho_{r,q}$, for purely real $\rho_A$. For purely imaginary  off-diagonal elements of $\rho_A$, we shall have $\rho_{q,r}=-\rho_{r,q}$, hence we need to choose $f(j)=f(d-j+1)$.
Then Eq.~\ref{eq:intermediatesigma} immediately simplifies to yield 
\begin{eqnarray}
\Tr[\rho^{\prime A} \mathcal{O}_A]&=&\sum_{p=1}^{d}\sum_{q,r}^d c_1  (U_A)_{p,q} \rho_{q, r} (U_A)^*_{r,p} \nonumber \\
&&=c_1 \Tr[\rho^{\prime A}]=c_1 
\label{eq:MUBvalue}
\end{eqnarray}

Thus we have proven that for any real  [imaginary] density matrix $\langle \mathcal{O}^{\prime}_A \otimes \mathbb{I} \rangle$ gives a constant value, $c_1$ depending on parameters of the operator chosen. If we perform the same unitary evolution on partition $B$ we will obtain a similar result for $\langle \mathbb{I} \otimes \mathcal{O}^{\prime}_B  \rangle$, say $c_2$. Then invoking the separability condition, $\rho=\sum_ip_i\rho_i^A \otimes \rho_i^B$ we can compute $\langle \mathcal{O}^{\prime}_A \otimes \mathcal{O}^{\prime}_B  \rangle$ to obtain $c_1 c_2$. Note that since the expectation values now depends on the operator being used and not the density matrix, for all values of index $i$ we shall obtain the same values.

Hence the connected correlator becomes 
\begin{eqnarray}
&&\langle \mathcal{O}^{\prime}_A \otimes \mathcal{O}^{\prime}_B  \rangle-\langle \mathcal{O}^{\prime}_A \otimes \mathbb{I} \rangle\langle \mathbb{I} \otimes \mathcal{O}^{\prime}_B  \rangle \nonumber \\
&=& \sum_{i=1}^d p_i c_1 c_2-(\sum_{i=1}^d p_i c_1) (\sum_{j=1}^d p_j c_2)\nonumber \\
&=& 0\label{eq:separable}
\end{eqnarray} 

Eq.~\ref{eq:separable} thus defines a  necessary condition for separability when the two conditions are fulfilled. Note that practically we can to choose $\mathcal{O}_A=\mathcal{O}_B$ to ensure that we measure only one observable. Hence, \\
\vspace{0.2 in}
{\centering\fbox{ \parbox{ \linewidth}{\textit{For a class of states, which, if separable, are only separable into purely real states [or with purely imaginary off-diagonal elements in $\rho$], a single measurement of the connected correlation in special mutually unbiased bases (MUBs) to the eigenbasis of the bipartite system is sufficient to detect entanglement. }}}}\\
\vspace{0.2 in}

 Thus with no prior knowledge of the state, a non-zero value
in our test would indicate it is not separable to completely real (or having completely imaginary off-diagonal
elements) states. This condition then immediately becomes useful to detect entanglement of rebits\cite{Caves2001,Wootters2012-WOOESI,Wootters_2014} which involves qubits in real Hilbert space. Previous experimental detection of entanglement of rebits\cite{PhysRevA.103.L040402} needed to take measures to ensure that the detection was truly in real space, but our method automatically ensures that entanglement, if detected, is true real space entanglement and thus simplifies the experimental setup considerably. In fact, it has been seen before that for classes of states bipartite entanglement and quantum simulation in qubit systems can also be achieved via system of rebits with an additional ancilla rebit\cite{PhysRevLett.102.020505,Koh_2018,Renou2021}.  Here our
method will be quite useful, but a complete mapping of our results for mixed states with the ancilla rebit method is beyond the scope of this work.

However, we can explicitly construct classes of density matrices (states) for which our result will hold for two qubits as follows.

The density matrix for a single qubit can be written in the most general form as $\frac{1}{2}(\mathbb{I}+v_j\sigma^j)$ where the index $j=1,2,3$, $\sigma^j$ denotes the Pauli matrices and $\vec{v}=v_j$ denotes a vector on the Bloch sphere. To ensure reality we need to restrict ourselves to the $XZ$ plane , i.e. choose $v_2=0$. This then allows to express one class of real separable two qubit density matrices as,
\begin{eqnarray}
    \rho_S&=&\sum_{i=1}^m p_i[\begin{pmatrix}
        \frac{1}{2}+\frac{\sqrt{1-r_i^2}}{2} & \frac{r_i}{2} \\
         \frac{r_i}{2} & \frac{1}{2}+\frac{\sqrt{1-r_i^2}}{2}  \\
    \end{pmatrix} \nonumber \\
    \otimes
     &&\begin{pmatrix}
        \frac{1}{2}+\frac{\sqrt{1-s_i^2}}{2} & \frac{s_i}{2} \\
         \frac{s_i}{2} & \frac{1}{2}+\frac{\sqrt{1-s_i^2}}{2}  \\
    \end{pmatrix}]
\end{eqnarray}
where ${r_i,s_i} \in [-1,1]$ and $\sum_{i=1}^m p_i=1$. $r_i$ and $s_i$ correspond to the $v_1$ element of the $i^{{\rm th}}$ density matrix in subsystems $A$ and $B$ respectively.  For simplicity let us choose $m=2$. We can then create a new class of states by mixing in entangling states with strength $\epsilon$ in the $\{\ket{00},\ket{11}\}$ plane of the form $\ket{\psi}=c \ket{00}+\sqrt{1-c^2}\ket{11}$. Subsequently, entanglement can be detected via $\mathcal{C}_2$ for the class of two qubit density matrices $\rho=(1-\epsilon) \rho_S+ \epsilon \ket{\psi}\bra{\psi}$ \footnote{We have chosen a simple form of the mixing state, more complex ones such as linear combinations of different such states are also valid parameterizations.} generated in this manner with the elements,
\begin{equation}
\begin{split}
    \rho_{11}&=\frac{1-\epsilon}{8}(a_1^{+}b_1^{+}+a_2^{+}b_2^{+})+c^2\epsilon \\
    \rho_{22}&=\frac{1-\epsilon}{8}(a_1^{+}b_1^{-}+a_2^{+}b_2^{-}) \\
    \rho_{33}&=\frac{1-\epsilon}{8}(a_1^{-}b_1^{+}+a_2^{-}b_2^{+}) \\
    \rho_{44}&=\frac{1-\epsilon}{8}(a_1^{-}b_1^{-}+a_2^{-}b_2^{-})+(1-c^2)\epsilon \\
    \rho_{12}&=\rho_{21}=\frac{1-\epsilon}{8}\sum_{i=1,2}s_i a_i^{+}\\
    \rho_{13}&=\rho_{31}=\frac{1-\epsilon}{8}\sum_{i=1,2}r_i b_i^{+}\\
    \rho_{14}&=\rho_{41}=\frac{1-\epsilon}{8}\sum_{i=1,2}r_i s_i +c \sqrt{1-c^2}\epsilon\\
    \rho_{23}&=\rho_{32}=\frac{1-\epsilon}{8}\sum_{i=1,2}r_i s_i \\
    \rho_{24}&=\rho_{42}=\frac{1-\epsilon}{8}\sum_{i=1,2}r_i b_i^{-} \\
    \rho_{34}&=\rho_{43}=\frac{1-\epsilon}{8}\sum_{i=1,2}s_i a_i^{-}  \\
\end{split}
\end{equation}
where $a_i^{\pm}=(1 \pm \sqrt{1-r_i^2})$,  $b_i^{\pm}=(1 \pm \sqrt{1-s_i^2})$ and $\epsilon, |c| \in [0,1]$ and $c \in \mathbb{R}$. One can verify this using exact diagonalization, by comparing with the known entanglement measure obtained from the positive partial transpose criterion, negativity $\mathcal{N}$, defined as
\begin{equation}
\mathcal{N} = \frac{||\rho^{\Gamma_A}||_1-1}{2},
\label{eq:Negativity}
\end{equation}
which returns values greater than $0$ iff $\epsilon>0, c \neq 0,1$. Here, $\rho^{\Gamma_A}$ represents the partial transpose with respect to subsystem $A$, and $||X||_1$ denotes the trace norm of operator $X$. This condition can be intuitively understood since the entangling perturbations are of strength $\epsilon$, and we have also verified this numerically (data not shown)\footnote{An analysis of exact eigenvalues is also possible since we need to solve quartic equations, but tedious.}. 
$\mathcal{C}_2$ with the choice of operator $\mathcal{O}_{A,B}=\sigma^z$ gives 
\begin{equation}
   \mathcal{C}_2=-2 \epsilon c \sqrt{1-c^2}
\end{equation} which agrees with $\mathcal{N}$ as an entanglement detection criterion. On the other hand $\mathcal{C}_1^{\epsilon=0}=\frac{1}{4}( \sqrt{1-r_1^2} \sqrt{1-s_1^2}-\sqrt{1-r_1^2} \sqrt{1-s_2^2}-\sqrt{1-r_2^2} \sqrt{1-s_1^2}+\sqrt{1-r_2^2} \sqrt{1-s_2^2}) \neq0$ which clearly does not correspond to the correct entanglement detection. This method can be immediately generalized to perturbations in other directions of the Hilbert space. Generalizations to more number of qubits will involve higher dimensional representations of the Bloch sphere.

 But one drawback of such a stringent condition lies in uncertainty of state preparations, thus in practice the value of zero for separable states becomes a strong lower bound instead of an exact result. However, in this framework we can make also an estimate of the maximal lower bound for separable states due a result of slight imperfections in state preparation. The result depends on choice of the observable.

Let us assume the separated density matrix of the subsystem has both real and imaginary parts $\rho_{q,r}=\rho^R_{q,r}+i \rho^I_{q,r}$. The we should recompute the derivation from the second step of Eq.~\eqref{eq:proof}  as,
\begin{eqnarray}
 &&\frac{1}{d}\sum_{p=1}^{d/2}\sum_{ q, r}^d  f(p) e^{-i \pi/2[(L_A-\phi_{p,q})-(L_A-\phi_{r,p})]} (\rho^R_{q,r}+i  \rho^I_{q,r})\nonumber \\
&&=\frac{1}{d}\sum_{p=1}^{d/2}\sum_{ q, r}^d  f(p) e^{-i \pi/2(\phi_{p,q}-\phi_{r,p})} \rho^R_{q,r} \nonumber \\
&&+\frac{1}{d}\sum_{p=1}^{d/2}\sum_{ \substack{ q, r \\ q \neq r}}^d  f(p) i e^{i \pi/2(\phi_{p,q}-\phi_{r,p})} \rho^I_{q,r}
\end{eqnarray}
We are going to try to find an upper bound of the value given by the extra contribution from the imaginary part. Noting that a similar term will also contribute from the first term of Eq.~\eqref{eq:intermediatesigma}, we can bound its contribution as,
\begin{eqnarray}
 &&|-\frac{2}{d}\sum_{p=1}^{d/2}\sum_{ q, r}^d  f(p) \sin( \pi/2(\phi_{p,q}-\phi_{r,p})) \rho^I_{q,r} | \nonumber \\ 
 &&=|-\frac{4}{d}\sum_{p=1}^{d/2} f(p) \sum_{ \substack{q, r \\ q<r}}^d \sin( \pi/2(\phi_{p,q}-\phi_{r,p})) \rho^I_{q,r} | \nonumber \\ 
\end{eqnarray}

Clearly the bound is dependent strongly on the distribution of the imaginary elements of the density matrix and only terms for which $(\phi_{p,q}-\phi_{r,p})=2k+1, k \in \mathbb{Z}$ contribute to the sum. Furthermore since $\sin((2k+1)\pi/2)=(-1)^k$, this sum contains terms of opposite signs. If  $|\rho^I_{q,r}| \sim O(\epsilon)$, as a rough estimate, we can approximate the sum as the expected deviation from the starting point of a one-dimensional classical random walk with step size $\epsilon$. Then we can define a upper bound by counting terms contributing to the sum, 
\begin{eqnarray*}
    &&|-\frac{4}{d}\sum_{p=1}^{d/2} f(p) \sum_{ \substack{q, r \\ q<r}}^d \sin( \pi/2(\phi_{p,q}-\phi_{r,p})) \rho^I_{q,r} | \nonumber \\ 
    &&\le\frac{4}{d}\sum_{p=1}^{d/2}| f(p)| \sum_{ \substack{q, r \\ q<r}}^d | \sin( \pi/2(\phi_{p,q}-\phi_{r,p})) \rho^I_{q,r} | \nonumber \\
    && \sim \frac{4 \epsilon}{d} \sqrt{\frac{d^2}{4}}\sum_{p=1}^{d/2}|f(p)|= |2 \epsilon \sum_{p=1}^{d/2}f(p)|
\end{eqnarray*}
ignoring some prefactors, since only $d^2/4$ terms from $\frac{d(d-1)}{2}$ terms will contribute. For large $d$, typically $\epsilon \sim O(1/d)$ for random states, then the bound becomes,
\begin{equation}
    \frac{2}{d}\sum_{p=1}^{d/2}|f(p)|
\end{equation}
From this naive computation we expect that the error vanishes in the thermodynamic limit. In fact, later in Example II where the purely real (imaginary) conditions are not satisfied and we have to define a cutoff based on numerical results, we actually show in Appendix~\ref{app:appD} the cutoff actually becomes smaller with system size.

 It is also worth noting that the observable $\mathcal{O}^{\prime}_A$, obtained via unitarily transforming $\mathcal{O}_A$ by $U_A$ that satisfies our purpose, is not unique. However $U_A$, which also serves as the eigenbasis to the new observable $\mathcal{O}^{\prime}_A$, cannot be any arbitrary complete set of MUBs to the computational basis. Only certain special rotations (unitary evolutions) serve to generate the correct eigenbasis mutually unbiased to the original basis to satisfy Eq. 13. We have not been able to find a general way to construct all such basis but in Appendix~\ref{app:appB}, we provide another explicit example of $U_A$, which although is not as local as the current example, it also works for a collection of qudits. Further note that if we were able to obtain $U_{A[B]}$ of the above form but with the additional property,
\begin{equation}
 \phi_{p,q}-\phi_{p,r}=2m,  \hspace{0.2 in} m \in \mathbb{Z}
 \label{eq:spcond}
\end{equation}
this technique would be applicable without the first condition. Unfortunately, this renders the matrix non-unitary and thus is an invalid choice. However with conserved local magnetization (the alternative condition), only specific elements of $U_A$ are responsible for the rotation. In this set-up, Eq.~\eqref{eq:spcond} is fulfilled, leading to applicability for generic density matrices.  We explain this aspect in details in the following.

\paragraph*{\textbf{Proof of condition (ii):}}First note that if indeed we obtain separable states with fixed total magnetization, then the magnetization of $\rho^A_i$ and $\rho^B_i$ needs to be fixed. However, this does not imply the subsystems have fixed magnetization, but just that the distribution of magnetization between the two subsystems become explicit. For example if we have a separable state with $\langle \sum\limits_{j=1}^L\sigma^z_j\rangle=1$, it could be written as,
\begin{eqnarray*}
    \rho&=&p_1 \rho^A_{1,\Sigma^A=1} \otimes  \rho^B_{1,\Sigma^B=0}
    +\\ && p_2 \rho^A_{2,\Sigma^A=-1} \otimes  \rho^B_{2,\Sigma^B=2}+ \hdots + \\ && p_N \rho^A_{N,\Sigma^A=l} \otimes  \rho^B_{N,\Sigma^B=1-l}
\end{eqnarray*}
where $\sum_{i=1}^N p_i=1,\Sigma^{A[B]}=\langle\sum_{j=1}^{L_{A[B]}}\sigma^z_j\rangle$, the subscripts denote the magnetization sector of each $\rho_i$ of the corresponding subsystem, and the values it can take are limited by the total number of sites in the subsystem ($l$ and $1-l$ are some allowed values within these limits). It can easily be seen that if each such $\rho_i^{A[B]}$ did not have a fixed magnetization, we would no longer be confined to the $\langle \sum\limits_{j=1}^L\sigma^z_j\rangle=1$ magnetization sector of $\rho$ of the full system. 

Let us revisit Eq.~\eqref{eq:proof} for such a situation. Because now non-zero elements of $\rho_i^{A[B]}$ are confined to a particular $\langle \sum\sigma^z\rangle$, only certain indices $q,r$ contributes to the sum--- the indices corresponding to the basis states with that particular $\langle \sum\sigma^z\rangle$. For example if we have $3$ sites and $\langle \sum\limits_{j=1}^3\sigma^z_j\rangle=1$ the relevant basis states would be $\ket{011}, \ket{101}, \ket{110}$, since in all the three cases $\langle \sum_{j=1}^3\sigma^z_j \rangle=\langle\sum_{j=1}^3  (2n_j-1) \rangle=1$, $n_j$ being the number of particles at the site $j$. We shall again drop the subscript $i$ rom $\rho_A$ in what follows. 

Since non-zero elements of $\rho^A$ are confined to certain indices, when we multiply by $U_A$, the column indices which are relevant will also be confined to those indices. Let us assume we have an element $U_{p,q_M}$ where $q_M$ denotes particular column indices corresponding to the conserved magnetization $M$. A similar situation will arise for the indices of $U_A^*$ of Eq.~\eqref{eq:proof} where the row (or column, due to $U^T=U$) index $r$ can take limited values $r_M$.

Now we shall show that under these restrictions $(\phi_{p,q_M}-\phi_{r_M,p})= (\phi_{p,q_M}-\phi_{p, r_M})=2m$. If we explicitly write out the computational basis, we realize that only two kinds of operations are possible to go from any $q_M$ to any $r_M$. (i) The trivial case of keeping the spin at a site constant. (ii) The non-trivial case of switching positions of spins between two sites. From the properties of $U$ it is easy to show that the second case introduces a factor of $a a^{-1}=1$ or $a^{\pm 2}$ (where $a=e^{-i \pi/2}$),(ignoring prefactors) since it involves two spin flips, while the first introduces a trivial $1$. Due to the tensor product structure of $U$, this means any the relevant indices are always connected by factors of the form $a^{2m}$ as $a^{\pm 2}$ and $1$ are the only operations allowed.

To take a concrete example, consider the case where $L=3,p=4, S_z=-1$. The allowed values of $q_M$ and $r_M$ are $2,3,5$ which denotes the computational states $\ket{001},\ket{010}, \ket{100}$. Since $p \in [1,d/2] \cap \mathbb{Z}$  we have chosen $p=4$ whose computational state is $\ket{011}$, for generality. Ignoring the unimportant prefactors, element $(4,2)$ is then given by $\bra{0}U_{L=1}\ket{0}\bra{1}U_{L=1}\ket{0}\bra{1}U_{L=1}\ket{1}=1\times a\times1=a$, where we have used Eq.~\eqref{eq:onesite}. Element $(4,5)$ is found as $\bra{0}U_{L=1}\ket{1}\bra{1}U_{L=1}\ket{0}\bra{1}U_{L=1}\ket{0}=a\times a\times a=a^3$. The factor of $a^2$ coming from the double spin flip is apparent. And finally element  
$(4,3)$ is $\bra{0}U_{L=1}\ket{0}\bra{1}U_{L=1}\ket{1}\bra{1}U_{L=1}\ket{0}=1\times 1\times a=a$.
In this case the flips in the first two site from the previous element gives a ratio of $\frac{\bra{0}U_{L=1}\ket{0}\bra{1}U_{L=1}\ket{1}}{\bra{0}U_{L=1}\ket{1}\bra{1}U_{L=1}\ket{0}}=a^{-2}$. Since $a=e^{i \pi/2}$, and the change $e^{i(\phi_{p,q_M}-\phi_{p, r_M})}$ is by a factor of $a^{2m}=e^{i m \pi}=e^{i (2m) \pi/2 }$ for $m=1$ or $m=0$, Eq.~\eqref{eq:spcond} is satisfied.  

To complete the proof, notice that in this case we can write $e^{-i \pi/2(\phi_{pq}-\phi_{rp})}=e^{-i m \pi}=e^{i m \pi}=e^{-i \pi/2(\phi_{rp}-\phi_{pq})}=e^{-i \pi/2[(L-\phi_{pq})-(L-\phi_{rp})]}$. Then Eq.~\eqref{eq:proof} follows without any assumption on $\rho$. Similar arguments can be provided for other locally conserved charges.

It is worthy to note that, an additional intuitive argument exists to prove the requirement of only one set of measurements to detect entanglement in generic density matrices with locally conserved charges. If a locally conserved charge is removed from a subsystem of a bipartite system, it has to necessarily appear in the other subsystem, i.e. the two halves are always correlated if one measures the conserved charge. Then we borrow the result from Eq.~\eqref{eq:Macconecond} and choose $\mathcal{O}_{A[B]}$ to be the conserved local charge, say for example the magnetization,  i.e. $\sum_{i=1}^{L_A} \langle \sigma^z_i\rangle+\sum_{i=L_A+1}^{L} \langle \sigma^z_i\rangle=M$, then by choosing $\mathcal{O_A}=\sum_{i=1}^{L_A} \sigma^z_i$ and $\mathcal{O_B}=\sum_{i=L_A+1}^{L} \sigma^z_i$, $|\mathcal{P}_\mathcal{O}|=1$ as the two observables are always completely linearly correlated (an analytic proof of this statement is provided in Appendix~\ref{app:appC}). This reduces Eq.~\eqref{eq:Macconecond} to,
\begin{equation}
    |\mathcal{P}_\mathcal{O^{\prime}}|>0
\end{equation}
 Since $\mathcal{P}_\mathcal{O^{\prime}}$ is just the rescaled connected correlation, a non-zero connected correlation in any MUB to the computational basis will indicate entanglement . Since the unitary evolution in Eq.~\eqref{eq:basis1} generates such a basis, we shall be able to detect entanglement between $A$ and $B$ by measuring correlation between the total magnetization of the subsystems 
after the evolution. Thus we have,
\begin{figure}
\centering
\includegraphics[width=0.9\columnwidth]{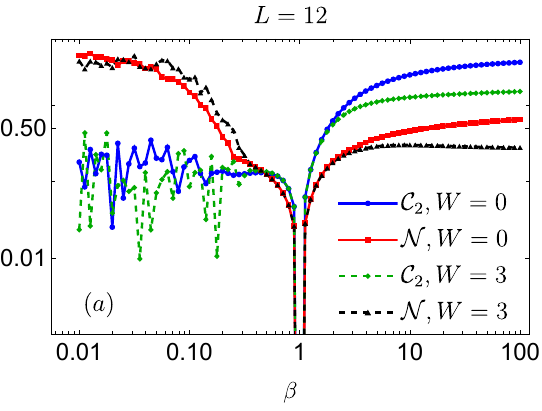}
\includegraphics[width=0.9\columnwidth]{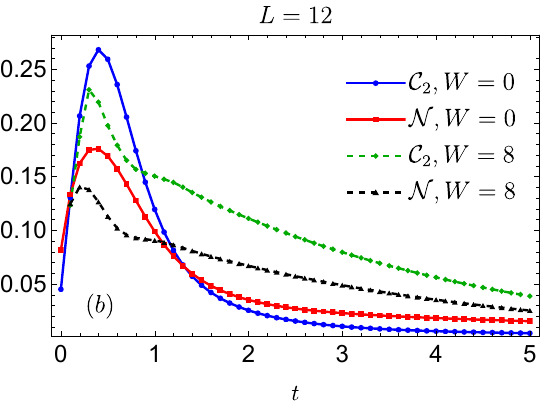}\\

\caption{ Comparison of behaviour of $\mathcal{C}_2$ [Eq.~\eqref{eq:C2}] with $\mathcal{N}$ [Eq.~\eqref{eq:Negativity}] for top: different values of $\beta$ for states defined in Eq.~\eqref{eq:thermal} of the model defined in Eq.~\eqref{eq:Heisen} for two different $W$ and bottom: for dissipative evolution via Eq.~\eqref{eq:lindblad} starting from  N\'eel state. $\mathcal{O}_A=\sum_{i=1}^{L_A}  \sigma^z_i$ and $\mathcal{O}_B=\sum_{i=L_A+1}^{L_B}\sigma^z_i$, $U_{A[B]}$ is given by Eq.~\eqref{eq:basis1}  }
\label{fig:fig1}
\end{figure}
\begin{figure*}
\centering
\includegraphics[width=0.9\columnwidth]{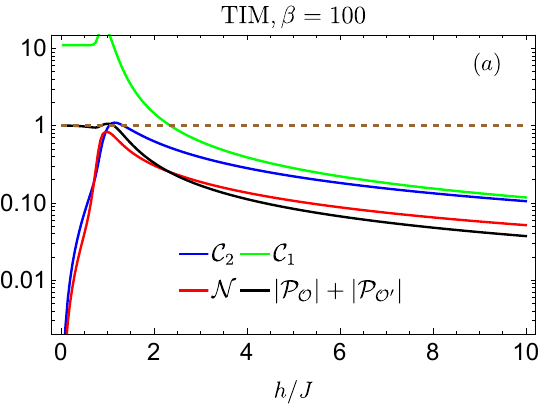}
\includegraphics[width=0.9\columnwidth]{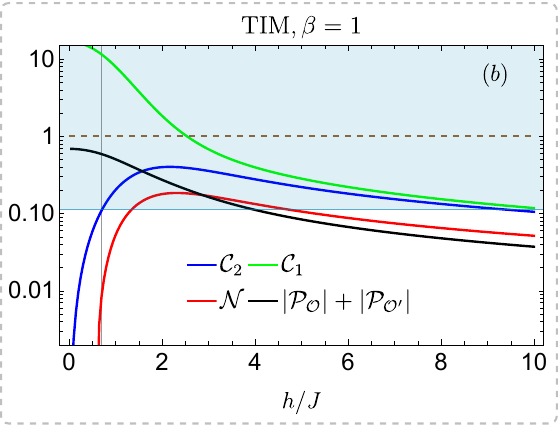}\\
\includegraphics[width=0.9\columnwidth]{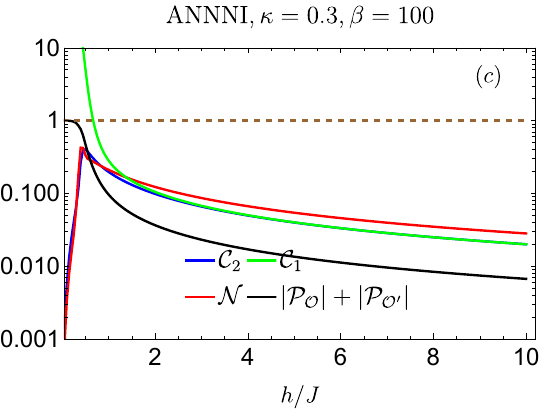}
\includegraphics[width=0.9\columnwidth]{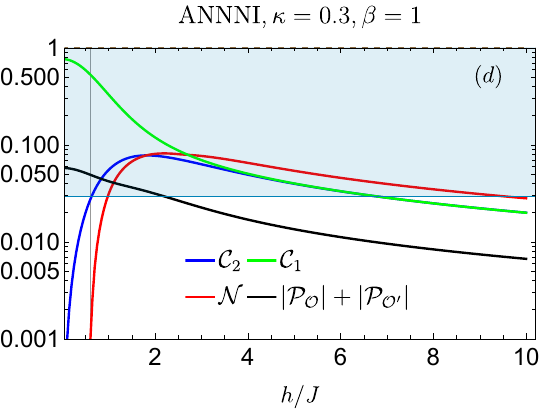}\\
\caption{Comparison of behaviour of $\mathcal{C}_1$ and $\mathcal{C}_2$ defined in Eqs.~\eqref{eq:C1},\eqref{eq:C2}, to  $\mathcal{N}$[Eq.~\eqref{eq:Negativity}] and the criterion in Eq.~\eqref{eq:Macconecond} for the model given in Eq.~\eqref{eq:ising}, for(a)  $\kappa=0$ at $\beta=1/T=100$, (b) $\kappa=0$ at $\beta=1$, (c) $\kappa=0.3$ at $\beta=100$ and (d) $\kappa=0.3$ at $\beta=1$. The system size is $L=12$, the blue shaded region in (b) and (d) shows the region where $\mathcal{C}_2$ is above the cut-off and entanglement detection is possible. The brown dashed line denotes the cut-off $1$ above with entanglement detection is possible via Eq.~\eqref{eq:Macconecond}. A small longitudinal field of strength $10^{-3}$ is added to break degeneracy.}
\label{fig:fig2}
\end{figure*}

\vspace{0.2 in}
{\centering\fbox{ \parbox{ \linewidth}{\textit{In presence of a locally conserved charge, a single measurement of connected correlation in mutually unbiased bases (MUBs) to the eigenbasis of the of the bipartite system is sufficient to detect entanglement of arbitrary states.}}}}\\
\vspace{0.2 in}


 \paragraph*{\textbf{Numerical tests:}}To test the efficacy of this technique in experimentally detecting entanglement of mixed states generated by generic Hamiltonians, we numerically compare $\mathcal{C}_2$ with a quantity originating from the PPT criterion\cite{PhysRevLett.77.1413,HORODECKI19961}, entanglement negativity\cite{PhysRevA.65.032314,PhysRevLett.95.090503,PhysRevB.81.064429,santos2011negativity,gray2019scale} $\mathcal{N}$.We provide two examples below, (i) Heisenberg models, which conserves magnetization, (ii) Ising models, which do not conserve magnetization. A third example of PXP model, a constrained model which does not conserve magnetization, is provided in Appendix~\ref{app:appE}.  In these examples, the quantum state whose entanglement we intend to measure is typically in thermal equilibrium with a corresponding initial Hamiltonian. Then to measure $\mathcal{C}_2$ which requires a specific time evolution of the state, we need to perform a `quantum quench', where we unitarily evolve the equilibrium state with a new effective Hamiltonian having a totally different eigenbasis which in our case is simply $\sum_j \sigma^x_j$. 

Let us take a brief detour to elaborate a bit more about quantum quench. When we evolve a quantum state by a Hamiltonian, we can take two extreme approaches. In one case we slowly change the Hamiltonian parameter such that during the evolution the state always remains the ground state of the changing Hamiltonian. This is adiabatic evolution and is typically utilized in quantum annealing. \cite{rajak2023quantum} However the other scenario is starting from the ground (eigen) state of a Hamiltonian we suddenly change the parameters of the Hamiltonian and then allow the system to relax with the new parameters. Clearly this induces non-adiabaticity in the system, and results in a completely different non-trivial evolution of the state. For example, if $\ket{\psi(0)}$ is the initial pure state we start from, since it is not an eigenstate of the new Hamiltonian, the time evolution can be expressed as,
\begin{eqnarray*}
  \ket{\psi(t)}=e^{-i H t} \ket{\psi(0)}=\sum_{j=1}^d e^{-i E_j t} \ket{E_j} \braket{E_j|\psi(0)}
\end{eqnarray*}
where $E_j$ and $\ket{E_j}$ are the eigenvalues and corresponding eigenvectors of the new Hamiltonian $H$. This non-trivial dynamics allows us to reach our desired MUB to measure $\mathcal{C}_2$. 
We shall see that while for the Heisenberg model it immediately detects entanglement, for the Ising models, additional information about the state is needed to detect entanglement.

\subparagraph{ Example I--- \textit{Heisenberg model with quasi periodic disorder:}}
The Heisenberg model with on-site quasiperiodic disorder is described by the following Hamiltonian\cite{doi:10.1073/pnas.1800589115},

\begin{equation}
H = -J \sum_k \bm{\sigma}_k \cdot \bm{\sigma}_{k+1} + \sum_k h_k \sigma_k^z.
\label{eq:Heisen}
\end{equation}

In this equation, $\bm{\sigma} = (\sigma^x, \sigma^y, \sigma^z)$ represents a vector of Pauli matrices. The term $h_k$ is the quasiperiodic component given by $h_k = W \cos(2 \pi \eta k)$, where $\eta = \frac{\sqrt{5}-1}{2}$ and $W$ represents the strength of the disorder. We take $J=1$. The inclusion of the quasiperiodic term allows for the study of a more general model, while eliminating the need for averaging over realizations that would be required with random disorder.

We consider computing the entanglement in the following two cases, (i)for the equilibrium density matrix at any temperature $T=1/\beta$, given by the Gibbs state,
\begin{equation}
\rho=\sum_{i=1}^{d^2}\frac{e^{-\beta \epsilon_i}}{\sum_{i=1}^{d^2}e^{-\beta \epsilon_i}} \ket{E_i} \bra{E_i}
\label{eq:thermal}
\end{equation} where $\epsilon_i$ denotes the eigenenergies and $\ket{E_i}$ denotes the corresponding eigenvectors; 
(ii) a time evolution in an open quantum system where the unitary evolution is governed via the Hamiltonian in Eq.~\eqref{eq:Heisen} and the influence of the environment is effectively included via the Lindblad equation of motion~\cite{Lindblad1976}:

\begin{equation}
\frac{d\rho}{dt} = i[H, \rho] + \sum_k ([L_k \rho, L_k^\dagger] + [L_k, \rho L_k^\dagger]),
\label{eq:lindblad}
\end{equation}

where $\rho$ represents the density matrix describing the system, and $H$ is the Hamiltonian from Eq.~\eqref{eq:Heisen}. The Lindblad operators $L_k$ capture the interactions between the system and the environment. We consider only on-site dephasing terms of strength $\gamma$, $L_k=\sqrt{\frac{\gamma}{2}}\sigma^z_k$, which causes the system to eventually evolve to the maximally mixed state, and ensures conservation of magnetization. We choose the initial state to be the N\'eel state\cite{PhysRevB.93.094205,PhysRevLett.116.160401}. We have selected these specific cases in order to examine and analyse mixed states, thereby avoiding the trivial scenario of pure state entanglement detection.  We restrict our system to the $\sum_{k=1}^L\sigma^z_k=0$ sector.

In Fig.\ref{fig:fig1}(a), we present the results for case (i). Notably, $\mathcal{C}_2$ (Eq.\eqref{eq:C2}) with observables $\mathcal{O}_{A[B]}=\sum_{i=1}^{L_{A[B]}} \sigma^z_i$, and unitary transformations $U_{A[B]}$ [Eq.\eqref{eq:basis1}], successfully detects entanglement in the same parameter regimes where $\mathcal{N}$ is greater than $0$. This correspondence demonstrates that our criterion of non-zero $\mathcal{C}_2$ aligns with the violation of the positive partial transpose (PPT) criterion for both instances of disorder considered. Additionally, it is worth noting that $\mathcal{C}_2$ exhibits a value of $0$ at $\beta=1$, consistent with separability according to the PPT criterion. Moreover, the {\em qualitative} agreement between $\mathcal{C}_2$ and $\mathcal{N}$ becomes more pronounced at higher $\beta$ values, corresponding to purer states.

In Fig.~\ref{fig:fig1}(b), where we examine case (ii), there is no separability during the time evolution. However, $\mathcal{C}_2$ exhibits the `entanglement barrier' feature previously observed in other measures of entanglement and also in $\mathcal{N}$ \cite{PRXQuantum.4.010318}. For larger disorder, we observe that the barrier shifts to lower time-scales due to the earlier decay of long-range correlations in such systems. Furthermore, the entanglement decay slows down on addition of disorder, i.e. disorder helps protect coherence during the transient dynamics \cite{ghoshPhysRevB.107.184303}. Thus, the measurement of $\mathcal{C}_2$ provides a reliable technique to detect entanglement. It is worth noting that in the presence of locally conserved charge, one can adjust the quench procedure to suit the experiment as long as an observable is generated whose eigenbasis is mutually unbiased with respect to the computational basis.
\subparagraph{ Example II--- \textit{Anisotropic next-nearest neighbour Transverse Ising model:}}
Next, consider the model Hamiltonian,
\begin{equation}
H=-J\sum_k \sigma^z_k \sigma^z_{k+1}+\kappa \sum_k \sigma^z_k \sigma^z_{k+2}-h \sum_k \sigma^x_k
\label{eq:ising}
\end{equation}
This is the non-integrable anisotropic next-nearest neighbour Transverse Ising (ANNNI) model, which reduces to the integrable Transverse Ising (TI) model when $\kappa=0$. Note that this model does not have a simple locally conserved charge such as magnetization, so quantities such as number entropy or particle fluctuations cannot detect entanglement. 

In the limit of $\beta \rightarrow \infty$, the Transverse Ising (TI) model undergoes a second-order phase transition at $h=h_c=1$ for $L \rightarrow \infty$. This phase transition is also observed in the anisotropic next-nearest neighbour Transverse Ising (ANNNI) model, but within a limited parameter range. As we gradually increase the parameter $\kappa$ in the ANNNI model, $h_c$ decreases, and at $\kappa=0.5$, the critical value becomes $h_c=0$.  \cite{PhysRevB.73.052402,Suzuki2013,PhysRevX.11.031062}. Once again, we investigate the behaviour of $\mathcal{C}_2$ in comparison to $\mathcal{N}$ in the thermal states of the models described by the ANNNI Hamiltonian. The thermal states are given by Eq.~\eqref{eq:thermal}, but with the Hamiltonian $H$ now specified by the ANNNI model.

From Fig.~\ref{fig:fig2}, it is evident that in all scenarios, the behaviour of $\mathcal{C}_2$ for the same observables as before is qualitatively similar to that of the  entanglement measure $\mathcal{N}$. However, the agreement of $\mathcal{C}_1$ is dependent on the chosen parameters.\footnote{Since we work with just the connected correlation functions we sometimes may obtain large values of $\mathcal{C}_1$ and $\mathcal{C_2}$. Because we are mostly concerned with detection and not measurement we deem rescaling unnecessary. Bounds can however be found depending on the system size an operators involved. For the transverse Ising model in our work where we use total magnetization operator, the peak occurs at $h=0$ and takes a value $L^2/4$. This can be seen by computing the Gibbs state in the presence of infinitesimal longitudinal field at small but finite temperature--- $\frac{1}{2}\ket{\uparrow\uparrow\hdots\uparrow}\bra{\uparrow\uparrow\hdots\uparrow}+\frac{1}{2}\ket{\downarrow\downarrow\hdots\downarrow}\bra{\downarrow\downarrow\hdots\downarrow}$. This state gives the maximum possible unnormalized connected correlation for this observable, and in fact $\mathcal{C}_2$ is loosely upper bounded by this value as well since we always rotate away from this special state.} Furthermore, the criteria in Eq.~\eqref{eq:Macconecond}, $|\mathcal{P}_\mathcal{O}|+|\mathcal{P}_\mathcal{O}^{\prime}| >1$, denoted by the brown dashed line, is not fulfilled almost everywhere, thus being a poor entanglement witness in this set-up. An improvement could potentially be achieved by increasing the number of mutually unbiased basis (MUB) measurements, although the analysis of this is beyond the scope of the present study. However, a complete agreement between $\mathcal{C}_2$ and $\mathcal{N}$ for all values of $h$ with $\kappa=0$ and $\kappa=0.3$ only occurs at large values of $\beta$, where the state becomes minimally mixed. 

For smaller $\beta$, there is a region where $\mathcal{N}=0$ while $\mathcal{C}_2$ has a small value. Here the system is separable via PPT criterion, but not to purely real states (or there exists bound entanglement).\footnote{It is not separable to states with purely imaginary off diagonal elements either as taking $\mathcal{O}=\sum \langle |\sigma_z|\rangle$ does not yield $\mathcal{C}_2=0$ in this region} However there still exists a qualitative similarity of $\mathcal{C}_2$ with $\mathcal{N}$, which is due to the stability of the correlator to perturbation with imaginary separable states. We can utilize this qualitative similarity to define a lower non-zero cut-off of $\mathcal{C}_2$ at each $\beta$ which can be used to detect entanglement. The cut-off can be found using numerical simulations for small size systems and extrapolated to larger sizes, due to weak dependence on system size (See Appendix~\ref{app:appD}) and thus provides a viable route to detect entanglement of states at thermal equilibrium. Note that an arbitrary MUB to computational basis does not yield qualitatively similar results to $\mathcal{N}$ in this case unlike the previous one.

The experimental realization of the Transverse Ising model is possible, for example, in quantum wires by tuning the chemical potential appropriately. \cite{Meyer_2009}. After letting the system reach thermal equilibrium, to perform the rotation one would need perform a quench by `pinching' between neighbouring electrons to ensure $J=\kappa=0$ and let the system evolve for time $t=\frac{\pi}{4 h}+2 k \pi$,for $k \in \mathbb{Z}$ and measure the total $\sigma_z$ correlations between the corresponding parts of the system. After the `pinch' the evolution occurs via an on-site Hamiltonian, hence the evolution does not generate entanglement. Then by using the thresholds for the specific temperature, we can detect entanglement.

\paragraph*{\textbf{Conclusion:}} In this work we have demonstrated through an analytic proof how measuring the one-connected correlation in the appropriate basis can provide valuable insights into the separability of the state, and effectively detect entanglement by leveraging specific properties of the state. This approach offers a substantial reduction in the experimental burden for a wide range of systems, as compared to previous correlation measurement methods that often require multiple basis measurements to achieve efficient detection. Moreover, our method presents an alternative means of detecting entanglement in charge-conserved systems.

We have showcased the immediate applicability of our technique in detecting entanglement in Heisenberg chains, where our technique provides an efficient entanglement detection pathway for both pure and mixed states and in quantum wires, where conventional correlation measurements face challenges in detecting entanglement. Although the detection in the latter case may not be immediate, the qualitative similarity between our correlator and the Negativity measure allows for the establishment of thresholds based on experimental conditions to effectively detect entanglement.  Additionally, we can also detect and quantify certain errors in full state preparation by computing $\mathcal{C}_2$ between two uncoupled thermal states which should be identically $0$ for perfectly prepared states. For example, lack of equilibriation will give us $\mathcal{C}_2 \neq 0$ as then we will have imaginary elements in the off-diagonal density matrix due to unitary time evolution.

Overall, our work provides a valuable approach for entanglement detection that offers both efficiency and applicability across various systems. It significantly simplifies possibilities of entanglement detection in experiments.



\begin{acknowledgements}
    We thank Felix Fritzsch, Daniele Toniolo, Madhumita Sarkar, Marko \v{Z}nidari\v{c}, Alexander Nico-Katz, Crispin Barnes and Debarshi Das for discussions and suggestions. We also thank three anonymous referees for their valuable comments which have helped improve our work considerably. The authors acknowledge the EPSRC grant Nonergodic quantum manipulation EP/R029075/1 for support.
\end{acknowledgements}

\bibliography{MUB} 

\begin{thebibliography}{66}%
\makeatletter
\providecommand \@ifxundefined [1]{%
 \@ifx{#1\undefined}
}%
\providecommand \@ifnum [1]{%
 \ifnum #1\expandafter \@firstoftwo
 \else \expandafter \@secondoftwo
 \fi
}%
\providecommand \@ifx [1]{%
 \ifx #1\expandafter \@firstoftwo
 \else \expandafter \@secondoftwo
 \fi
}%
\providecommand \natexlab [1]{#1}%
\providecommand \enquote  [1]{``#1''}%
\providecommand \bibnamefont  [1]{#1}%
\providecommand \bibfnamefont [1]{#1}%
\providecommand \citenamefont [1]{#1}%
\providecommand \href@noop [0]{\@secondoftwo}%
\providecommand \href [0]{\begingroup \@sanitize@url \@href}%
\providecommand \@href[1]{\@@startlink{#1}\@@href}%
\providecommand \@@href[1]{\endgroup#1\@@endlink}%
\providecommand \@sanitize@url [0]{\catcode `\\12\catcode `\$12\catcode
  `\&12\catcode `\#12\catcode `\^12\catcode `\_12\catcode `\%12\relax}%
\providecommand \@@startlink[1]{}%
\providecommand \@@endlink[0]{}%
\providecommand \url  [0]{\begingroup\@sanitize@url \@url }%
\providecommand \@url [1]{\endgroup\@href {#1}{\urlprefix }}%
\providecommand \urlprefix  [0]{URL }%
\providecommand \Eprint [0]{\href }%
\providecommand \doibase [0]{https://doi.org/}%
\providecommand \selectlanguage [0]{\@gobble}%
\providecommand \bibinfo  [0]{\@secondoftwo}%
\providecommand \bibfield  [0]{\@secondoftwo}%
\providecommand \translation [1]{[#1]}%
\providecommand \BibitemOpen [0]{}%
\providecommand \bibitemStop [0]{}%
\providecommand \bibitemNoStop [0]{.\EOS\space}%
\providecommand \EOS [0]{\spacefactor3000\relax}%
\providecommand \BibitemShut  [1]{\csname bibitem#1\endcsname}%
\let\auto@bib@innerbib\@empty
\bibitem [{\citenamefont {Islam}\ \emph {et~al.}(2015)\citenamefont {Islam},
  \citenamefont {Ma}, \citenamefont {Preiss}, \citenamefont {Eric~Tai},
  \citenamefont {Lukin}, \citenamefont {Rispoli},\ and\ \citenamefont
  {Greiner}}]{Islam2015}%
  \BibitemOpen
  \bibfield  {author} {\bibinfo {author} {\bibfnamefont {R.}~\bibnamefont
  {Islam}}, \bibinfo {author} {\bibfnamefont {R.}~\bibnamefont {Ma}}, \bibinfo
  {author} {\bibfnamefont {P.~M.}\ \bibnamefont {Preiss}}, \bibinfo {author}
  {\bibfnamefont {M.}~\bibnamefont {Eric~Tai}}, \bibinfo {author}
  {\bibfnamefont {A.}~\bibnamefont {Lukin}}, \bibinfo {author} {\bibfnamefont
  {M.}~\bibnamefont {Rispoli}},\ and\ \bibinfo {author} {\bibfnamefont
  {M.}~\bibnamefont {Greiner}},\ }\bibfield  {title} {\bibinfo {title}
  {Measuring entanglement entropy in a quantum many-body system},\ }\href
  {https://doi.org/10.1038/nature15750} {\bibfield  {journal} {\bibinfo
  {journal} {Nature}\ }\textbf {\bibinfo {volume} {528}},\ \bibinfo {pages}
  {77} (\bibinfo {year} {2015})}\BibitemShut {NoStop}%
\bibitem [{\citenamefont {Brydges}\ \emph {et~al.}(2019)\citenamefont
  {Brydges}, \citenamefont {Elben}, \citenamefont {Jurcevic}, \citenamefont
  {Vermersch}, \citenamefont {Maier}, \citenamefont {Lanyon}, \citenamefont
  {Zoller}, \citenamefont {Blatt},\ and\ \citenamefont
  {Roos}}]{zollerscience.aau4963}%
  \BibitemOpen
  \bibfield  {author} {\bibinfo {author} {\bibfnamefont {T.}~\bibnamefont
  {Brydges}}, \bibinfo {author} {\bibfnamefont {A.}~\bibnamefont {Elben}},
  \bibinfo {author} {\bibfnamefont {P.}~\bibnamefont {Jurcevic}}, \bibinfo
  {author} {\bibfnamefont {B.}~\bibnamefont {Vermersch}}, \bibinfo {author}
  {\bibfnamefont {C.}~\bibnamefont {Maier}}, \bibinfo {author} {\bibfnamefont
  {B.~P.}\ \bibnamefont {Lanyon}}, \bibinfo {author} {\bibfnamefont
  {P.}~\bibnamefont {Zoller}}, \bibinfo {author} {\bibfnamefont
  {R.}~\bibnamefont {Blatt}},\ and\ \bibinfo {author} {\bibfnamefont {C.~F.}\
  \bibnamefont {Roos}},\ }\bibfield  {title} {\bibinfo {title} {Probing
  r\'{e}nyi entanglement entropy via randomized measurements},\ }\href
  {https://doi.org/10.1126/science.aau4963} {\bibfield  {journal} {\bibinfo
  {journal} {Science}\ }\textbf {\bibinfo {volume} {364}},\ \bibinfo {pages}
  {260} (\bibinfo {year} {2019})},\ \Eprint
  {https://arxiv.org/abs/https://www.science.org/doi/pdf/10.1126/science.aau4963}
  {https://www.science.org/doi/pdf/10.1126/science.aau4963} \BibitemShut
  {NoStop}%
\bibitem [{\citenamefont {Liu}\ \emph {et~al.}(2022{\natexlab{a}})\citenamefont
  {Liu}, \citenamefont {Liu}, \citenamefont {Chen},\ and\ \citenamefont
  {Ma}}]{PhysRevLett.129.230503}%
  \BibitemOpen
  \bibfield  {author} {\bibinfo {author} {\bibfnamefont {P.}~\bibnamefont
  {Liu}}, \bibinfo {author} {\bibfnamefont {Z.}~\bibnamefont {Liu}}, \bibinfo
  {author} {\bibfnamefont {S.}~\bibnamefont {Chen}},\ and\ \bibinfo {author}
  {\bibfnamefont {X.}~\bibnamefont {Ma}},\ }\bibfield  {title} {\bibinfo
  {title} {Fundamental limitation on the detectability of entanglement},\
  }\href {https://doi.org/10.1103/PhysRevLett.129.230503} {\bibfield  {journal}
  {\bibinfo  {journal} {Phys. Rev. Lett.}\ }\textbf {\bibinfo {volume} {129}},\
  \bibinfo {pages} {230503} (\bibinfo {year} {2022}{\natexlab{a}})}\BibitemShut
  {NoStop}%
\bibitem [{\citenamefont {Dowling}\ \emph {et~al.}(2004)\citenamefont
  {Dowling}, \citenamefont {Doherty},\ and\ \citenamefont
  {Bartlett}}]{PhysRevA.70.062113}%
  \BibitemOpen
  \bibfield  {author} {\bibinfo {author} {\bibfnamefont {M.~R.}\ \bibnamefont
  {Dowling}}, \bibinfo {author} {\bibfnamefont {A.~C.}\ \bibnamefont
  {Doherty}},\ and\ \bibinfo {author} {\bibfnamefont {S.~D.}\ \bibnamefont
  {Bartlett}},\ }\bibfield  {title} {\bibinfo {title} {Energy as an
  entanglement witness for quantum many-body systems},\ }\href
  {https://doi.org/10.1103/PhysRevA.70.062113} {\bibfield  {journal} {\bibinfo
  {journal} {Phys. Rev. A}\ }\textbf {\bibinfo {volume} {70}},\ \bibinfo
  {pages} {062113} (\bibinfo {year} {2004})}\BibitemShut {NoStop}%
\bibitem [{\citenamefont {Fr\'erot}\ \emph {et~al.}(2022)\citenamefont
  {Fr\'erot}, \citenamefont {Baccari},\ and\ \citenamefont
  {Ac\'{\i}n}}]{PRXQuantum.3.010342}%
  \BibitemOpen
  \bibfield  {author} {\bibinfo {author} {\bibfnamefont {I.}~\bibnamefont
  {Fr\'erot}}, \bibinfo {author} {\bibfnamefont {F.}~\bibnamefont {Baccari}},\
  and\ \bibinfo {author} {\bibfnamefont {A.}~\bibnamefont {Ac\'{\i}n}},\
  }\bibfield  {title} {\bibinfo {title} {Unveiling quantum entanglement in
  many-body systems from partial information},\ }\href
  {https://doi.org/10.1103/PRXQuantum.3.010342} {\bibfield  {journal} {\bibinfo
   {journal} {PRX Quantum}\ }\textbf {\bibinfo {volume} {3}},\ \bibinfo {pages}
  {010342} (\bibinfo {year} {2022})}\BibitemShut {NoStop}%
\bibitem [{\citenamefont {Chiara}\ and\ \citenamefont
  {Sanpera}(2018)}]{DeChiara_2018}%
  \BibitemOpen
  \bibfield  {author} {\bibinfo {author} {\bibfnamefont {G.~D.}\ \bibnamefont
  {Chiara}}\ and\ \bibinfo {author} {\bibfnamefont {A.}~\bibnamefont
  {Sanpera}},\ }\bibfield  {title} {\bibinfo {title} {Genuine quantum
  correlations in quantum many-body systems: a review of recent progress},\
  }\href {https://doi.org/10.1088/1361-6633/aabf61} {\bibfield  {journal}
  {\bibinfo  {journal} {Reports on Progress in Physics}\ }\textbf {\bibinfo
  {volume} {81}},\ \bibinfo {pages} {074002} (\bibinfo {year}
  {2018})}\BibitemShut {NoStop}%
\bibitem [{\citenamefont {Liu}\ \emph {et~al.}(2022{\natexlab{b}})\citenamefont
  {Liu}, \citenamefont {Tang}, \citenamefont {Dai}, \citenamefont {Liu},
  \citenamefont {Chen},\ and\ \citenamefont {Ma}}]{PhysRevLett.129.260501}%
  \BibitemOpen
  \bibfield  {author} {\bibinfo {author} {\bibfnamefont {Z.}~\bibnamefont
  {Liu}}, \bibinfo {author} {\bibfnamefont {Y.}~\bibnamefont {Tang}}, \bibinfo
  {author} {\bibfnamefont {H.}~\bibnamefont {Dai}}, \bibinfo {author}
  {\bibfnamefont {P.}~\bibnamefont {Liu}}, \bibinfo {author} {\bibfnamefont
  {S.}~\bibnamefont {Chen}},\ and\ \bibinfo {author} {\bibfnamefont
  {X.}~\bibnamefont {Ma}},\ }\bibfield  {title} {\bibinfo {title} {Detecting
  entanglement in quantum many-body systems via permutation moments},\ }\href
  {https://doi.org/10.1103/PhysRevLett.129.260501} {\bibfield  {journal}
  {\bibinfo  {journal} {Phys. Rev. Lett.}\ }\textbf {\bibinfo {volume} {129}},\
  \bibinfo {pages} {260501} (\bibinfo {year} {2022}{\natexlab{b}})}\BibitemShut
  {NoStop}%
\bibitem [{\citenamefont {Neven}\ \emph {et~al.}(2021)\citenamefont {Neven},
  \citenamefont {Carrasco}, \citenamefont {Vitale}, \citenamefont {Kokail},
  \citenamefont {Elben}, \citenamefont {Dalmonte}, \citenamefont {Calabrese},
  \citenamefont {Zoller}, \citenamefont {Vermersch}, \citenamefont {Kueng},\
  and\ \citenamefont {Kraus}}]{Neven2021}%
  \BibitemOpen
  \bibfield  {author} {\bibinfo {author} {\bibfnamefont {A.}~\bibnamefont
  {Neven}}, \bibinfo {author} {\bibfnamefont {J.}~\bibnamefont {Carrasco}},
  \bibinfo {author} {\bibfnamefont {V.}~\bibnamefont {Vitale}}, \bibinfo
  {author} {\bibfnamefont {C.}~\bibnamefont {Kokail}}, \bibinfo {author}
  {\bibfnamefont {A.}~\bibnamefont {Elben}}, \bibinfo {author} {\bibfnamefont
  {M.}~\bibnamefont {Dalmonte}}, \bibinfo {author} {\bibfnamefont
  {P.}~\bibnamefont {Calabrese}}, \bibinfo {author} {\bibfnamefont
  {P.}~\bibnamefont {Zoller}}, \bibinfo {author} {\bibfnamefont
  {B.}~\bibnamefont {Vermersch}}, \bibinfo {author} {\bibfnamefont
  {R.}~\bibnamefont {Kueng}},\ and\ \bibinfo {author} {\bibfnamefont
  {B.}~\bibnamefont {Kraus}},\ }\bibfield  {title} {\bibinfo {title}
  {Symmetry-resolved entanglement detection using partial transpose moments},\
  }\href {https://doi.org/10.1038/s41534-021-00487-y} {\bibfield  {journal}
  {\bibinfo  {journal} {npj Quantum Information}\ }\textbf {\bibinfo {volume}
  {7}},\ \bibinfo {pages} {152} (\bibinfo {year} {2021})}\BibitemShut {NoStop}%
\bibitem [{\citenamefont {Vitale}\ \emph {et~al.}(2022)\citenamefont {Vitale},
  \citenamefont {Elben}, \citenamefont {Kueng}, \citenamefont {Neven},
  \citenamefont {Carrasco}, \citenamefont {Kraus}, \citenamefont {Zoller},
  \citenamefont {Calabrese}, \citenamefont {Vermersch},\ and\ \citenamefont
  {Dalmonte}}]{10.21468/SciPostPhys.12.3.106}%
  \BibitemOpen
  \bibfield  {author} {\bibinfo {author} {\bibfnamefont {V.}~\bibnamefont
  {Vitale}}, \bibinfo {author} {\bibfnamefont {A.}~\bibnamefont {Elben}},
  \bibinfo {author} {\bibfnamefont {R.}~\bibnamefont {Kueng}}, \bibinfo
  {author} {\bibfnamefont {A.}~\bibnamefont {Neven}}, \bibinfo {author}
  {\bibfnamefont {J.}~\bibnamefont {Carrasco}}, \bibinfo {author}
  {\bibfnamefont {B.}~\bibnamefont {Kraus}}, \bibinfo {author} {\bibfnamefont
  {P.}~\bibnamefont {Zoller}}, \bibinfo {author} {\bibfnamefont
  {P.}~\bibnamefont {Calabrese}}, \bibinfo {author} {\bibfnamefont
  {B.}~\bibnamefont {Vermersch}},\ and\ \bibinfo {author} {\bibfnamefont
  {M.}~\bibnamefont {Dalmonte}},\ }\bibfield  {title} {\bibinfo {title}
  {{Symmetry-resolved dynamical purification in synthetic quantum matter}},\
  }\href {https://doi.org/10.21468/SciPostPhys.12.3.106} {\bibfield  {journal}
  {\bibinfo  {journal} {SciPost Phys.}\ }\textbf {\bibinfo {volume} {12}},\
  \bibinfo {pages} {106} (\bibinfo {year} {2022})}\BibitemShut {NoStop}%
\bibitem [{\citenamefont {Zhou}\ \emph {et~al.}(2020)\citenamefont {Zhou},
  \citenamefont {Zeng},\ and\ \citenamefont {Liu}}]{PhysRevLett.125.200502}%
  \BibitemOpen
  \bibfield  {author} {\bibinfo {author} {\bibfnamefont {Y.}~\bibnamefont
  {Zhou}}, \bibinfo {author} {\bibfnamefont {P.}~\bibnamefont {Zeng}},\ and\
  \bibinfo {author} {\bibfnamefont {Z.}~\bibnamefont {Liu}},\ }\bibfield
  {title} {\bibinfo {title} {Single-copies estimation of entanglement
  negativity},\ }\href {https://doi.org/10.1103/PhysRevLett.125.200502}
  {\bibfield  {journal} {\bibinfo  {journal} {Phys. Rev. Lett.}\ }\textbf
  {\bibinfo {volume} {125}},\ \bibinfo {pages} {200502} (\bibinfo {year}
  {2020})}\BibitemShut {NoStop}%
\bibitem [{\citenamefont {Gray}\ \emph {et~al.}(2018)\citenamefont {Gray},
  \citenamefont {Banchi}, \citenamefont {Bayat},\ and\ \citenamefont
  {Bose}}]{gray2018machine}%
  \BibitemOpen
  \bibfield  {author} {\bibinfo {author} {\bibfnamefont {J.}~\bibnamefont
  {Gray}}, \bibinfo {author} {\bibfnamefont {L.}~\bibnamefont {Banchi}},
  \bibinfo {author} {\bibfnamefont {A.}~\bibnamefont {Bayat}},\ and\ \bibinfo
  {author} {\bibfnamefont {S.}~\bibnamefont {Bose}},\ }\bibfield  {title}
  {\bibinfo {title} {Machine-learning-assisted many-body entanglement
  measurement},\ }\href {https://doi.org/10.1103/PhysRevLett.121.150503}
  {\bibfield  {journal} {\bibinfo  {journal} {Phys. Rev. Lett.}\ }\textbf
  {\bibinfo {volume} {121}},\ \bibinfo {pages} {150503} (\bibinfo {year}
  {2018})}\BibitemShut {NoStop}%
\bibitem [{\citenamefont {Banchi}\ \emph {et~al.}(2016)\citenamefont {Banchi},
  \citenamefont {Bayat},\ and\ \citenamefont {Bose}}]{banchi2016entanglement}%
  \BibitemOpen
  \bibfield  {author} {\bibinfo {author} {\bibfnamefont {L.}~\bibnamefont
  {Banchi}}, \bibinfo {author} {\bibfnamefont {A.}~\bibnamefont {Bayat}},\ and\
  \bibinfo {author} {\bibfnamefont {S.}~\bibnamefont {Bose}},\ }\bibfield
  {title} {\bibinfo {title} {Entanglement entropy scaling in solid-state spin
  arrays via capacitance measurements},\ }\href
  {https://doi.org/10.1103/PhysRevB.94.241117} {\bibfield  {journal} {\bibinfo
  {journal} {Phys. Rev. B}\ }\textbf {\bibinfo {volume} {94}},\ \bibinfo
  {pages} {241117} (\bibinfo {year} {2016})}\BibitemShut {NoStop}%
\bibitem [{\citenamefont {Squillante}\ \emph {et~al.}(2023)\citenamefont
  {Squillante}, \citenamefont {Ricco}, \citenamefont {Ukpong}, \citenamefont
  {Lagos-Monaco}, \citenamefont {Seridonio},\ and\ \citenamefont
  {de~Souza}}]{squillante2023gr}%
  \BibitemOpen
  \bibfield  {author} {\bibinfo {author} {\bibfnamefont {L.}~\bibnamefont
  {Squillante}}, \bibinfo {author} {\bibfnamefont {L.~S.}\ \bibnamefont
  {Ricco}}, \bibinfo {author} {\bibfnamefont {A.~M.}\ \bibnamefont {Ukpong}},
  \bibinfo {author} {\bibfnamefont {R.~E.}\ \bibnamefont {Lagos-Monaco}},
  \bibinfo {author} {\bibfnamefont {A.~C.}\ \bibnamefont {Seridonio}},\ and\
  \bibinfo {author} {\bibfnamefont {M.}~\bibnamefont {de~Souza}},\ }\bibfield
  {title} {\bibinfo {title} {Gr\" uneisen parameter as an entanglement
  compass},\ }\href@noop {} {\bibfield  {journal} {\bibinfo  {journal} {arXiv
  preprint arXiv:2306.00566}\ } (\bibinfo {year} {2023})}\BibitemShut {NoStop}%
\bibitem [{\citenamefont {Verstraete}\ \emph {et~al.}(2004)\citenamefont
  {Verstraete}, \citenamefont {Popp},\ and\ \citenamefont {Cirac}}]{Cirac1}%
  \BibitemOpen
  \bibfield  {author} {\bibinfo {author} {\bibfnamefont {F.}~\bibnamefont
  {Verstraete}}, \bibinfo {author} {\bibfnamefont {M.}~\bibnamefont {Popp}},\
  and\ \bibinfo {author} {\bibfnamefont {J.~I.}\ \bibnamefont {Cirac}},\
  }\bibfield  {title} {\bibinfo {title} {Entanglement versus correlations in
  spin systems},\ }\href {https://doi.org/10.1103/PhysRevLett.92.027901}
  {\bibfield  {journal} {\bibinfo  {journal} {Phys. Rev. Lett.}\ }\textbf
  {\bibinfo {volume} {92}},\ \bibinfo {pages} {027901} (\bibinfo {year}
  {2004})}\BibitemShut {NoStop}%
\bibitem [{\citenamefont {Rappoport}\ \emph {et~al.}(2007)\citenamefont
  {Rappoport}, \citenamefont {Ghivelder}, \citenamefont {Fernandes},
  \citenamefont {Guimar\~aes},\ and\ \citenamefont
  {Continentino}}]{PhysRevB.75.054422}%
  \BibitemOpen
  \bibfield  {author} {\bibinfo {author} {\bibfnamefont {T.~G.}\ \bibnamefont
  {Rappoport}}, \bibinfo {author} {\bibfnamefont {L.}~\bibnamefont
  {Ghivelder}}, \bibinfo {author} {\bibfnamefont {J.~C.}\ \bibnamefont
  {Fernandes}}, \bibinfo {author} {\bibfnamefont {R.~B.}\ \bibnamefont
  {Guimar\~aes}},\ and\ \bibinfo {author} {\bibfnamefont {M.~A.}\ \bibnamefont
  {Continentino}},\ }\bibfield  {title} {\bibinfo {title} {Experimental
  observation of quantum entanglement in low-dimensional spin systems},\ }\href
  {https://doi.org/10.1103/PhysRevB.75.054422} {\bibfield  {journal} {\bibinfo
  {journal} {Phys. Rev. B}\ }\textbf {\bibinfo {volume} {75}},\ \bibinfo
  {pages} {054422} (\bibinfo {year} {2007})}\BibitemShut {NoStop}%
\bibitem [{\citenamefont {Igl\'oi}\ and\ \citenamefont
  {T\'oth}(2023)}]{PhysRevResearch.5.013158}%
  \BibitemOpen
  \bibfield  {author} {\bibinfo {author} {\bibfnamefont {F.}~\bibnamefont
  {Igl\'oi}}\ and\ \bibinfo {author} {\bibfnamefont {G.}~\bibnamefont
  {T\'oth}},\ }\bibfield  {title} {\bibinfo {title} {Entanglement witnesses in
  the $xy$ chain: Thermal equilibrium and postquench nonequilibrium states},\
  }\href {https://doi.org/10.1103/PhysRevResearch.5.013158} {\bibfield
  {journal} {\bibinfo  {journal} {Phys. Rev. Res.}\ }\textbf {\bibinfo {volume}
  {5}},\ \bibinfo {pages} {013158} (\bibinfo {year} {2023})}\BibitemShut
  {NoStop}%
\bibitem [{\citenamefont {T\'oth}(2005)}]{PhysRevA.71.010301}%
  \BibitemOpen
  \bibfield  {author} {\bibinfo {author} {\bibfnamefont {G.}~\bibnamefont
  {T\'oth}},\ }\bibfield  {title} {\bibinfo {title} {Entanglement witnesses in
  spin models},\ }\href {https://doi.org/10.1103/PhysRevA.71.010301} {\bibfield
   {journal} {\bibinfo  {journal} {Phys. Rev. A}\ }\textbf {\bibinfo {volume}
  {71}},\ \bibinfo {pages} {010301} (\bibinfo {year} {2005})}\BibitemShut
  {NoStop}%
\bibitem [{\citenamefont {Wu}\ \emph {et~al.}(2005)\citenamefont {Wu},
  \citenamefont {Bandyopadhyay}, \citenamefont {Sarandy},\ and\ \citenamefont
  {Lidar}}]{PhysRevA.72.032309}%
  \BibitemOpen
  \bibfield  {author} {\bibinfo {author} {\bibfnamefont {L.-A.}\ \bibnamefont
  {Wu}}, \bibinfo {author} {\bibfnamefont {S.}~\bibnamefont {Bandyopadhyay}},
  \bibinfo {author} {\bibfnamefont {M.~S.}\ \bibnamefont {Sarandy}},\ and\
  \bibinfo {author} {\bibfnamefont {D.~A.}\ \bibnamefont {Lidar}},\ }\bibfield
  {title} {\bibinfo {title} {Entanglement observables and witnesses for
  interacting quantum spin systems},\ }\href
  {https://doi.org/10.1103/PhysRevA.72.032309} {\bibfield  {journal} {\bibinfo
  {journal} {Phys. Rev. A}\ }\textbf {\bibinfo {volume} {72}},\ \bibinfo
  {pages} {032309} (\bibinfo {year} {2005})}\BibitemShut {NoStop}%
\bibitem [{\citenamefont {Song}\ \emph {et~al.}(2010)\citenamefont {Song},
  \citenamefont {Rachel},\ and\ \citenamefont {Le~Hur}}]{PhysRevB.82.012405}%
  \BibitemOpen
  \bibfield  {author} {\bibinfo {author} {\bibfnamefont {H.~F.}\ \bibnamefont
  {Song}}, \bibinfo {author} {\bibfnamefont {S.}~\bibnamefont {Rachel}},\ and\
  \bibinfo {author} {\bibfnamefont {K.}~\bibnamefont {Le~Hur}},\ }\bibfield
  {title} {\bibinfo {title} {General relation between entanglement and
  fluctuations in one dimension},\ }\href
  {https://doi.org/10.1103/PhysRevB.82.012405} {\bibfield  {journal} {\bibinfo
  {journal} {Phys. Rev. B}\ }\textbf {\bibinfo {volume} {82}},\ \bibinfo
  {pages} {012405} (\bibinfo {year} {2010})}\BibitemShut {NoStop}%
\bibitem [{\citenamefont {Petrescu}\ \emph {et~al.}(2014)\citenamefont
  {Petrescu}, \citenamefont {Song}, \citenamefont {Rachel}, \citenamefont
  {Ristivojevic}, \citenamefont {Flindt}, \citenamefont {Laflorencie},
  \citenamefont {Klich}, \citenamefont {Regnault},\ and\ \citenamefont
  {Hur}}]{Petrescu_2014}%
  \BibitemOpen
  \bibfield  {author} {\bibinfo {author} {\bibfnamefont {A.}~\bibnamefont
  {Petrescu}}, \bibinfo {author} {\bibfnamefont {H.~F.}\ \bibnamefont {Song}},
  \bibinfo {author} {\bibfnamefont {S.}~\bibnamefont {Rachel}}, \bibinfo
  {author} {\bibfnamefont {Z.}~\bibnamefont {Ristivojevic}}, \bibinfo {author}
  {\bibfnamefont {C.}~\bibnamefont {Flindt}}, \bibinfo {author} {\bibfnamefont
  {N.}~\bibnamefont {Laflorencie}}, \bibinfo {author} {\bibfnamefont
  {I.}~\bibnamefont {Klich}}, \bibinfo {author} {\bibfnamefont
  {N.}~\bibnamefont {Regnault}},\ and\ \bibinfo {author} {\bibfnamefont
  {K.~L.}\ \bibnamefont {Hur}},\ }\bibfield  {title} {\bibinfo {title}
  {Fluctuations and entanglement spectrum in quantum hall states},\ }\href
  {https://doi.org/10.1088/1742-5468/2014/10/P10005} {\bibfield  {journal}
  {\bibinfo  {journal} {Journal of Statistical Mechanics: Theory and
  Experiment}\ }\textbf {\bibinfo {volume} {2014}},\ \bibinfo {pages} {P10005}
  (\bibinfo {year} {2014})}\BibitemShut {NoStop}%
\bibitem [{\citenamefont {Song}\ \emph {et~al.}(2012)\citenamefont {Song},
  \citenamefont {Rachel}, \citenamefont {Flindt}, \citenamefont {Klich},
  \citenamefont {Laflorencie},\ and\ \citenamefont
  {Le~Hur}}]{PhysRevB.85.035409}%
  \BibitemOpen
  \bibfield  {author} {\bibinfo {author} {\bibfnamefont {H.~F.}\ \bibnamefont
  {Song}}, \bibinfo {author} {\bibfnamefont {S.}~\bibnamefont {Rachel}},
  \bibinfo {author} {\bibfnamefont {C.}~\bibnamefont {Flindt}}, \bibinfo
  {author} {\bibfnamefont {I.}~\bibnamefont {Klich}}, \bibinfo {author}
  {\bibfnamefont {N.}~\bibnamefont {Laflorencie}},\ and\ \bibinfo {author}
  {\bibfnamefont {K.}~\bibnamefont {Le~Hur}},\ }\bibfield  {title} {\bibinfo
  {title} {Bipartite fluctuations as a probe of many-body entanglement},\
  }\href {https://doi.org/10.1103/PhysRevB.85.035409} {\bibfield  {journal}
  {\bibinfo  {journal} {Phys. Rev. B}\ }\textbf {\bibinfo {volume} {85}},\
  \bibinfo {pages} {035409} (\bibinfo {year} {2012})}\BibitemShut {NoStop}%
\bibitem [{\citenamefont {Song}\ \emph {et~al.}(2011)\citenamefont {Song},
  \citenamefont {Flindt}, \citenamefont {Rachel}, \citenamefont {Klich},\ and\
  \citenamefont {Le~Hur}}]{PhysRevB.83.161408}%
  \BibitemOpen
  \bibfield  {author} {\bibinfo {author} {\bibfnamefont {H.~F.}\ \bibnamefont
  {Song}}, \bibinfo {author} {\bibfnamefont {C.}~\bibnamefont {Flindt}},
  \bibinfo {author} {\bibfnamefont {S.}~\bibnamefont {Rachel}}, \bibinfo
  {author} {\bibfnamefont {I.}~\bibnamefont {Klich}},\ and\ \bibinfo {author}
  {\bibfnamefont {K.}~\bibnamefont {Le~Hur}},\ }\bibfield  {title} {\bibinfo
  {title} {Entanglement entropy from charge statistics: Exact relations for
  noninteracting many-body systems},\ }\href
  {https://doi.org/10.1103/PhysRevB.83.161408} {\bibfield  {journal} {\bibinfo
  {journal} {Phys. Rev. B}\ }\textbf {\bibinfo {volume} {83}},\ \bibinfo
  {pages} {161408} (\bibinfo {year} {2011})}\BibitemShut {NoStop}%
\bibitem [{\citenamefont {Oshima}\ and\ \citenamefont
  {Fuji}(2023)}]{PhysRevB.107.014308}%
  \BibitemOpen
  \bibfield  {author} {\bibinfo {author} {\bibfnamefont {H.}~\bibnamefont
  {Oshima}}\ and\ \bibinfo {author} {\bibfnamefont {Y.}~\bibnamefont {Fuji}},\
  }\bibfield  {title} {\bibinfo {title} {Charge fluctuation and charge-resolved
  entanglement in a monitored quantum circuit with $u(1)$ symmetry},\ }\href
  {https://doi.org/10.1103/PhysRevB.107.014308} {\bibfield  {journal} {\bibinfo
   {journal} {Phys. Rev. B}\ }\textbf {\bibinfo {volume} {107}},\ \bibinfo
  {pages} {014308} (\bibinfo {year} {2023})}\BibitemShut {NoStop}%
\bibitem [{\citenamefont {Macieszczak}\ \emph {et~al.}(2019)\citenamefont
  {Macieszczak}, \citenamefont {Levi}, \citenamefont {Macr\`{\i}},
  \citenamefont {Lesanovsky},\ and\ \citenamefont
  {Garrahan}}]{PhysRevA.99.052354}%
  \BibitemOpen
  \bibfield  {author} {\bibinfo {author} {\bibfnamefont {K.}~\bibnamefont
  {Macieszczak}}, \bibinfo {author} {\bibfnamefont {E.}~\bibnamefont {Levi}},
  \bibinfo {author} {\bibfnamefont {T.}~\bibnamefont {Macr\`{\i}}}, \bibinfo
  {author} {\bibfnamefont {I.}~\bibnamefont {Lesanovsky}},\ and\ \bibinfo
  {author} {\bibfnamefont {J.~P.}\ \bibnamefont {Garrahan}},\ }\bibfield
  {title} {\bibinfo {title} {Coherence, entanglement, and quantumness in closed
  and open systems with conserved charge, with an application to many-body
  localization},\ }\href {https://doi.org/10.1103/PhysRevA.99.052354}
  {\bibfield  {journal} {\bibinfo  {journal} {Phys. Rev. A}\ }\textbf {\bibinfo
  {volume} {99}},\ \bibinfo {pages} {052354} (\bibinfo {year}
  {2019})}\BibitemShut {NoStop}%
\bibitem [{\citenamefont {Lukin}\ \emph {et~al.}(2019)\citenamefont {Lukin},
  \citenamefont {Rispoli}, \citenamefont {Schittko}, \citenamefont {Tai},
  \citenamefont {Kaufman}, \citenamefont {Choi}, \citenamefont {Khemani},
  \citenamefont {Léonard},\ and\ \citenamefont
  {Greiner}}]{Lukinscience.aau0818}%
  \BibitemOpen
  \bibfield  {author} {\bibinfo {author} {\bibfnamefont {A.}~\bibnamefont
  {Lukin}}, \bibinfo {author} {\bibfnamefont {M.}~\bibnamefont {Rispoli}},
  \bibinfo {author} {\bibfnamefont {R.}~\bibnamefont {Schittko}}, \bibinfo
  {author} {\bibfnamefont {M.~E.}\ \bibnamefont {Tai}}, \bibinfo {author}
  {\bibfnamefont {A.~M.}\ \bibnamefont {Kaufman}}, \bibinfo {author}
  {\bibfnamefont {S.}~\bibnamefont {Choi}}, \bibinfo {author} {\bibfnamefont
  {V.}~\bibnamefont {Khemani}}, \bibinfo {author} {\bibfnamefont
  {J.}~\bibnamefont {Léonard}},\ and\ \bibinfo {author} {\bibfnamefont
  {M.}~\bibnamefont {Greiner}},\ }\bibfield  {title} {\bibinfo {title} {Probing
  entanglement in a many-body localized system},\ }\href
  {https://doi.org/10.1126/science.aau0818} {\bibfield  {journal} {\bibinfo
  {journal} {Science}\ }\textbf {\bibinfo {volume} {364}},\ \bibinfo {pages}
  {256} (\bibinfo {year} {2019})}\BibitemShut {NoStop}%
\bibitem [{\citenamefont {Kiefer-Emmanouilidis}\ \emph
  {et~al.}(2020)\citenamefont {Kiefer-Emmanouilidis}, \citenamefont {Unanyan},
  \citenamefont {Fleischhauer},\ and\ \citenamefont
  {Sirker}}]{PhysRevLett.124.243601}%
  \BibitemOpen
  \bibfield  {author} {\bibinfo {author} {\bibfnamefont {M.}~\bibnamefont
  {Kiefer-Emmanouilidis}}, \bibinfo {author} {\bibfnamefont {R.}~\bibnamefont
  {Unanyan}}, \bibinfo {author} {\bibfnamefont {M.}~\bibnamefont
  {Fleischhauer}},\ and\ \bibinfo {author} {\bibfnamefont {J.}~\bibnamefont
  {Sirker}},\ }\bibfield  {title} {\bibinfo {title} {Evidence for unbounded
  growth of the number entropy in many-body localized phases},\ }\href
  {https://doi.org/10.1103/PhysRevLett.124.243601} {\bibfield  {journal}
  {\bibinfo  {journal} {Phys. Rev. Lett.}\ }\textbf {\bibinfo {volume} {124}},\
  \bibinfo {pages} {243601} (\bibinfo {year} {2020})}\BibitemShut {NoStop}%
\bibitem [{\citenamefont {Ghosh}\ and\ \citenamefont {\ifmmode \check{Z}\else
  \v{Z}\fi{}nidari\ifmmode~\check{c}\else
  \v{c}\fi{}}(2022)}]{PhysRevB.105.144203}%
  \BibitemOpen
  \bibfield  {author} {\bibinfo {author} {\bibfnamefont {R.}~\bibnamefont
  {Ghosh}}\ and\ \bibinfo {author} {\bibfnamefont {M.}~\bibnamefont {\ifmmode
  \check{Z}\else \v{Z}\fi{}nidari\ifmmode~\check{c}\else \v{c}\fi{}}},\
  }\bibfield  {title} {\bibinfo {title} {Resonance-induced growth of number
  entropy in strongly disordered systems},\ }\href
  {https://doi.org/10.1103/PhysRevB.105.144203} {\bibfield  {journal} {\bibinfo
   {journal} {Phys. Rev. B}\ }\textbf {\bibinfo {volume} {105}},\ \bibinfo
  {pages} {144203} (\bibinfo {year} {2022})}\BibitemShut {NoStop}%
\bibitem [{\citenamefont {Han}\ \emph {et~al.}(2023)\citenamefont {Han},
  \citenamefont {Meir},\ and\ \citenamefont {Sela}}]{PhysRevLett.130.136201}%
  \BibitemOpen
  \bibfield  {author} {\bibinfo {author} {\bibfnamefont {C.}~\bibnamefont
  {Han}}, \bibinfo {author} {\bibfnamefont {Y.}~\bibnamefont {Meir}},\ and\
  \bibinfo {author} {\bibfnamefont {E.}~\bibnamefont {Sela}},\ }\bibfield
  {title} {\bibinfo {title} {Realistic protocol to measure entanglement at
  finite temperatures},\ }\href
  {https://doi.org/10.1103/PhysRevLett.130.136201} {\bibfield  {journal}
  {\bibinfo  {journal} {Phys. Rev. Lett.}\ }\textbf {\bibinfo {volume} {130}},\
  \bibinfo {pages} {136201} (\bibinfo {year} {2023})}\BibitemShut {NoStop}%
\bibitem [{\citenamefont {Jebarathinam}\ \emph {et~al.}(2020)\citenamefont
  {Jebarathinam}, \citenamefont {Home},\ and\ \citenamefont
  {Sinha}}]{PhysRevA.101.022112}%
  \BibitemOpen
  \bibfield  {author} {\bibinfo {author} {\bibfnamefont {C.}~\bibnamefont
  {Jebarathinam}}, \bibinfo {author} {\bibfnamefont {D.}~\bibnamefont {Home}},\
  and\ \bibinfo {author} {\bibfnamefont {U.}~\bibnamefont {Sinha}},\ }\bibfield
   {title} {\bibinfo {title} {Pearson correlation coefficient as a measure for
  certifying and quantifying high-dimensional entanglement},\ }\href
  {https://doi.org/10.1103/PhysRevA.101.022112} {\bibfield  {journal} {\bibinfo
   {journal} {Phys. Rev. A}\ }\textbf {\bibinfo {volume} {101}},\ \bibinfo
  {pages} {022112} (\bibinfo {year} {2020})}\BibitemShut {NoStop}%
\bibitem [{\citenamefont {Spengler}\ \emph {et~al.}(2012)\citenamefont
  {Spengler}, \citenamefont {Huber}, \citenamefont {Brierley}, \citenamefont
  {Adaktylos},\ and\ \citenamefont {Hiesmayr}}]{PhysRevA.86.022311}%
  \BibitemOpen
  \bibfield  {author} {\bibinfo {author} {\bibfnamefont {C.}~\bibnamefont
  {Spengler}}, \bibinfo {author} {\bibfnamefont {M.}~\bibnamefont {Huber}},
  \bibinfo {author} {\bibfnamefont {S.}~\bibnamefont {Brierley}}, \bibinfo
  {author} {\bibfnamefont {T.}~\bibnamefont {Adaktylos}},\ and\ \bibinfo
  {author} {\bibfnamefont {B.~C.}\ \bibnamefont {Hiesmayr}},\ }\bibfield
  {title} {\bibinfo {title} {Entanglement detection via mutually unbiased
  bases},\ }\href {https://doi.org/10.1103/PhysRevA.86.022311} {\bibfield
  {journal} {\bibinfo  {journal} {Phys. Rev. A}\ }\textbf {\bibinfo {volume}
  {86}},\ \bibinfo {pages} {022311} (\bibinfo {year} {2012})}\BibitemShut
  {NoStop}%
\bibitem [{\citenamefont {Maccone}\ \emph {et~al.}(2015)\citenamefont
  {Maccone}, \citenamefont {Bru\ss{}},\ and\ \citenamefont
  {Macchiavello}}]{Maccone1}%
  \BibitemOpen
  \bibfield  {author} {\bibinfo {author} {\bibfnamefont {L.}~\bibnamefont
  {Maccone}}, \bibinfo {author} {\bibfnamefont {D.}~\bibnamefont {Bru\ss{}}},\
  and\ \bibinfo {author} {\bibfnamefont {C.}~\bibnamefont {Macchiavello}},\
  }\bibfield  {title} {\bibinfo {title} {Complementarity and correlations},\
  }\href {https://doi.org/10.1103/PhysRevLett.114.130401} {\bibfield  {journal}
  {\bibinfo  {journal} {Phys. Rev. Lett.}\ }\textbf {\bibinfo {volume} {114}},\
  \bibinfo {pages} {130401} (\bibinfo {year} {2015})}\BibitemShut {NoStop}%
\bibitem [{\citenamefont {Sauerwein}\ \emph {et~al.}(2017)\citenamefont
  {Sauerwein}, \citenamefont {Macchiavello}, \citenamefont {Maccone},\ and\
  \citenamefont {Kraus}}]{Macconemulti}%
  \BibitemOpen
  \bibfield  {author} {\bibinfo {author} {\bibfnamefont {D.}~\bibnamefont
  {Sauerwein}}, \bibinfo {author} {\bibfnamefont {C.}~\bibnamefont
  {Macchiavello}}, \bibinfo {author} {\bibfnamefont {L.}~\bibnamefont
  {Maccone}},\ and\ \bibinfo {author} {\bibfnamefont {B.}~\bibnamefont
  {Kraus}},\ }\bibfield  {title} {\bibinfo {title} {Multipartite correlations
  in mutually unbiased bases},\ }\href
  {https://doi.org/10.1103/PhysRevA.95.042315} {\bibfield  {journal} {\bibinfo
  {journal} {Phys. Rev. A}\ }\textbf {\bibinfo {volume} {95}},\ \bibinfo
  {pages} {042315} (\bibinfo {year} {2017})}\BibitemShut {NoStop}%
\bibitem [{\citenamefont {Erker}\ \emph {et~al.}(2017)\citenamefont {Erker},
  \citenamefont {Krenn},\ and\ \citenamefont
  {Huber}}]{Erker2017quantifyinghigh}%
  \BibitemOpen
  \bibfield  {author} {\bibinfo {author} {\bibfnamefont {P.}~\bibnamefont
  {Erker}}, \bibinfo {author} {\bibfnamefont {M.}~\bibnamefont {Krenn}},\ and\
  \bibinfo {author} {\bibfnamefont {M.}~\bibnamefont {Huber}},\ }\bibfield
  {title} {\bibinfo {title} {Quantifying high dimensional entanglement with two
  mutually unbiased bases},\ }\href {https://doi.org/10.22331/q-2017-07-28-22}
  {\bibfield  {journal} {\bibinfo  {journal} {{Quantum}}\ }\textbf {\bibinfo
  {volume} {1}},\ \bibinfo {pages} {22} (\bibinfo {year} {2017})}\BibitemShut
  {NoStop}%
\bibitem [{\citenamefont {Hiesmayr}\ \emph {et~al.}(2021)\citenamefont
  {Hiesmayr}, \citenamefont {McNulty}, \citenamefont {Baek}, \citenamefont
  {Roy}, \citenamefont {Bae},\ and\ \citenamefont
  {Chruściński}}]{Hiesmayr_2021}%
  \BibitemOpen
  \bibfield  {author} {\bibinfo {author} {\bibfnamefont {B.~C.}\ \bibnamefont
  {Hiesmayr}}, \bibinfo {author} {\bibfnamefont {D.}~\bibnamefont {McNulty}},
  \bibinfo {author} {\bibfnamefont {S.}~\bibnamefont {Baek}}, \bibinfo {author}
  {\bibfnamefont {S.~S.}\ \bibnamefont {Roy}}, \bibinfo {author} {\bibfnamefont
  {J.}~\bibnamefont {Bae}},\ and\ \bibinfo {author} {\bibfnamefont
  {D.}~\bibnamefont {Chruściński}},\ }\bibfield  {title} {\bibinfo {title}
  {Detecting entanglement can be more effective with inequivalent mutually
  unbiased bases},\ }\href {https://doi.org/10.1088/1367-2630/ac20ea}
  {\bibfield  {journal} {\bibinfo  {journal} {New Journal of Physics}\ }\textbf
  {\bibinfo {volume} {23}},\ \bibinfo {pages} {093018} (\bibinfo {year}
  {2021})}\BibitemShut {NoStop}%
\bibitem [{\citenamefont {Sadana}\ \emph {et~al.}(2022)\citenamefont {Sadana},
  \citenamefont {Kanjilal}, \citenamefont {Home},\ and\ \citenamefont
  {Sinha}}]{sadana2022relating}%
  \BibitemOpen
  \bibfield  {author} {\bibinfo {author} {\bibfnamefont {S.}~\bibnamefont
  {Sadana}}, \bibinfo {author} {\bibfnamefont {S.}~\bibnamefont {Kanjilal}},
  \bibinfo {author} {\bibfnamefont {D.}~\bibnamefont {Home}},\ and\ \bibinfo
  {author} {\bibfnamefont {U.}~\bibnamefont {Sinha}},\ }\href@noop {} {\bibinfo
  {title} {Relating an entanglement measure with statistical correlators for
  two-qudit mixed states using only a pair of complementary observables}}
  (\bibinfo {year} {2022}),\ \Eprint {https://arxiv.org/abs/2201.06188}
  {arXiv:2201.06188 [quant-ph]} \BibitemShut {NoStop}%
\bibitem [{Note1()}]{Note1}%
  \BibitemOpen
  \bibinfo {note} {There are several methods already available for entanglement
  detection in this case,but mostly focusing on pure states. \cite
  {Petrescu_2014,PhysRevB.85.035409,PhysRevB.83.161408,PhysRevB.107.014308,PhysRevA.99.052354,Lukinscience.aau0818,PhysRevLett.124.243601,PhysRevB.105.144203,PhysRevLett.130.136201}.
  However this method provides a new correlation witness from the point of view
  of MUBs for all kinds of states.}\BibitemShut {Stop}%
\bibitem [{Note2()}]{Note2}%
  \BibitemOpen
  \bibinfo {note} {This means that if we want to measure bipartite entanglement
  not in real space but any other space, for example in momentum space, we have
  to choose the observable to be diagonal in the computational basis of that
  space.}\BibitemShut {Stop}%
\bibitem [{\citenamefont {Caves}\ \emph {et~al.}(2001)\citenamefont {Caves},
  \citenamefont {Fuchs},\ and\ \citenamefont {Rungta}}]{Caves2001}%
  \BibitemOpen
  \bibfield  {author} {\bibinfo {author} {\bibfnamefont {C.~M.}\ \bibnamefont
  {Caves}}, \bibinfo {author} {\bibfnamefont {C.~A.}\ \bibnamefont {Fuchs}},\
  and\ \bibinfo {author} {\bibfnamefont {P.}~\bibnamefont {Rungta}},\
  }\bibfield  {title} {\bibinfo {title} {Entanglement of formation of an
  arbitrary state of two rebits},\ }\href
  {https://doi.org/10.1023/A:1012215309321} {\bibfield  {journal} {\bibinfo
  {journal} {Foundations of Physics Letters}\ }\textbf {\bibinfo {volume}
  {14}},\ \bibinfo {pages} {199} (\bibinfo {year} {2001})}\BibitemShut
  {NoStop}%
\bibitem [{\citenamefont {Wootters}(2012)}]{Wootters2012-WOOESI}%
  \BibitemOpen
  \bibfield  {author} {\bibinfo {author} {\bibfnamefont {W.~K.}\ \bibnamefont
  {Wootters}},\ }\bibfield  {title} {\bibinfo {title} {Entanglement sharing in
  real-vector-space quantum theory},\ }\href
  {https://doi.org/10.1007/s10701-010-9488-1} {\bibfield  {journal} {\bibinfo
  {journal} {Foundations of Physics}\ }\textbf {\bibinfo {volume} {42}},\
  \bibinfo {pages} {19} (\bibinfo {year} {2012})}\BibitemShut {NoStop}%
\bibitem [{\citenamefont {Wootters}(2014)}]{Wootters_2014}%
  \BibitemOpen
  \bibfield  {author} {\bibinfo {author} {\bibfnamefont {W.~K.}\ \bibnamefont
  {Wootters}},\ }\bibfield  {title} {\bibinfo {title} {The rebit three-tangle
  and its relation to two-qubit entanglement},\ }\href
  {https://doi.org/10.1088/1751-8113/47/42/424037} {\bibfield  {journal}
  {\bibinfo  {journal} {Journal of Physics A: Mathematical and Theoretical}\
  }\textbf {\bibinfo {volume} {47}},\ \bibinfo {pages} {424037} (\bibinfo
  {year} {2014})}\BibitemShut {NoStop}%
\bibitem [{\citenamefont {Prasannan}\ \emph {et~al.}(2021)\citenamefont
  {Prasannan}, \citenamefont {De}, \citenamefont {Barkhofen}, \citenamefont
  {Brecht}, \citenamefont {Silberhorn},\ and\ \citenamefont
  {Sperling}}]{PhysRevA.103.L040402}%
  \BibitemOpen
  \bibfield  {author} {\bibinfo {author} {\bibfnamefont {N.}~\bibnamefont
  {Prasannan}}, \bibinfo {author} {\bibfnamefont {S.}~\bibnamefont {De}},
  \bibinfo {author} {\bibfnamefont {S.}~\bibnamefont {Barkhofen}}, \bibinfo
  {author} {\bibfnamefont {B.}~\bibnamefont {Brecht}}, \bibinfo {author}
  {\bibfnamefont {C.}~\bibnamefont {Silberhorn}},\ and\ \bibinfo {author}
  {\bibfnamefont {J.}~\bibnamefont {Sperling}},\ }\bibfield  {title} {\bibinfo
  {title} {Experimental entanglement characterization of two-rebit states},\
  }\href {https://doi.org/10.1103/PhysRevA.103.L040402} {\bibfield  {journal}
  {\bibinfo  {journal} {Phys. Rev. A}\ }\textbf {\bibinfo {volume} {103}},\
  \bibinfo {pages} {L040402} (\bibinfo {year} {2021})}\BibitemShut {NoStop}%
\bibitem [{\citenamefont {McKague}\ \emph {et~al.}(2009)\citenamefont
  {McKague}, \citenamefont {Mosca},\ and\ \citenamefont
  {Gisin}}]{PhysRevLett.102.020505}%
  \BibitemOpen
  \bibfield  {author} {\bibinfo {author} {\bibfnamefont {M.}~\bibnamefont
  {McKague}}, \bibinfo {author} {\bibfnamefont {M.}~\bibnamefont {Mosca}},\
  and\ \bibinfo {author} {\bibfnamefont {N.}~\bibnamefont {Gisin}},\ }\bibfield
   {title} {\bibinfo {title} {Simulating quantum systems using real hilbert
  spaces},\ }\href {https://doi.org/10.1103/PhysRevLett.102.020505} {\bibfield
  {journal} {\bibinfo  {journal} {Phys. Rev. Lett.}\ }\textbf {\bibinfo
  {volume} {102}},\ \bibinfo {pages} {020505} (\bibinfo {year}
  {2009})}\BibitemShut {NoStop}%
\bibitem [{\citenamefont {Koh}\ \emph {et~al.}(2018)\citenamefont {Koh},
  \citenamefont {Niu},\ and\ \citenamefont {Yoder}}]{Koh_2018}%
  \BibitemOpen
  \bibfield  {author} {\bibinfo {author} {\bibfnamefont {D.~E.}\ \bibnamefont
  {Koh}}, \bibinfo {author} {\bibfnamefont {M.~Y.}\ \bibnamefont {Niu}},\ and\
  \bibinfo {author} {\bibfnamefont {T.~J.}\ \bibnamefont {Yoder}},\ }\bibfield
  {title} {\bibinfo {title} {Quantum simulation from the bottom up: the case of
  rebits},\ }\href {https://doi.org/10.1088/1751-8121/aab9c4} {\bibfield
  {journal} {\bibinfo  {journal} {Journal of Physics A: Mathematical and
  Theoretical}\ }\textbf {\bibinfo {volume} {51}},\ \bibinfo {pages} {195302}
  (\bibinfo {year} {2018})}\BibitemShut {NoStop}%
\bibitem [{\citenamefont {Renou}\ \emph {et~al.}(2021)\citenamefont {Renou},
  \citenamefont {Trillo}, \citenamefont {Weilenmann}, \citenamefont {Le},
  \citenamefont {Tavakoli}, \citenamefont {Gisin}, \citenamefont {Ac{\'i}n},\
  and\ \citenamefont {Navascu{\'e}s}}]{Renou2021}%
  \BibitemOpen
  \bibfield  {author} {\bibinfo {author} {\bibfnamefont {M.-O.}\ \bibnamefont
  {Renou}}, \bibinfo {author} {\bibfnamefont {D.}~\bibnamefont {Trillo}},
  \bibinfo {author} {\bibfnamefont {M.}~\bibnamefont {Weilenmann}}, \bibinfo
  {author} {\bibfnamefont {T.~P.}\ \bibnamefont {Le}}, \bibinfo {author}
  {\bibfnamefont {A.}~\bibnamefont {Tavakoli}}, \bibinfo {author}
  {\bibfnamefont {N.}~\bibnamefont {Gisin}}, \bibinfo {author} {\bibfnamefont
  {A.}~\bibnamefont {Ac{\'i}n}},\ and\ \bibinfo {author} {\bibfnamefont
  {M.}~\bibnamefont {Navascu{\'e}s}},\ }\bibfield  {title} {\bibinfo {title}
  {Quantum theory based on real numbers can be experimentally falsified},\
  }\href {https://doi.org/10.1038/s41586-021-04160-4} {\bibfield  {journal}
  {\bibinfo  {journal} {Nature}\ }\textbf {\bibinfo {volume} {600}},\ \bibinfo
  {pages} {625} (\bibinfo {year} {2021})}\BibitemShut {NoStop}%
\bibitem [{Note3()}]{Note3}%
  \BibitemOpen
  \bibinfo {note} {We have chosen a simple form of the mixing state, more
  complex ones such as linear combinations of different such states are also
  valid parameterizations.}\BibitemShut {Stop}%
\bibitem [{Note4()}]{Note4}%
  \BibitemOpen
  \bibinfo {note} {An analysis of exact eigenvalues is also possible since we
  need to solve quartic equations, but tedious.}\BibitemShut {Stop}%
\bibitem [{\citenamefont {Peres}(1996)}]{PhysRevLett.77.1413}%
  \BibitemOpen
  \bibfield  {author} {\bibinfo {author} {\bibfnamefont {A.}~\bibnamefont
  {Peres}},\ }\bibfield  {title} {\bibinfo {title} {Separability criterion for
  density matrices},\ }\href {https://doi.org/10.1103/PhysRevLett.77.1413}
  {\bibfield  {journal} {\bibinfo  {journal} {Phys. Rev. Lett.}\ }\textbf
  {\bibinfo {volume} {77}},\ \bibinfo {pages} {1413} (\bibinfo {year}
  {1996})}\BibitemShut {NoStop}%
\bibitem [{\citenamefont {Horodecki}\ \emph {et~al.}(1996)\citenamefont
  {Horodecki}, \citenamefont {Horodecki},\ and\ \citenamefont
  {Horodecki}}]{HORODECKI19961}%
  \BibitemOpen
  \bibfield  {author} {\bibinfo {author} {\bibfnamefont {M.}~\bibnamefont
  {Horodecki}}, \bibinfo {author} {\bibfnamefont {P.}~\bibnamefont
  {Horodecki}},\ and\ \bibinfo {author} {\bibfnamefont {R.}~\bibnamefont
  {Horodecki}},\ }\bibfield  {title} {\bibinfo {title} {Separability of mixed
  states: necessary and sufficient conditions},\ }\href
  {https://doi.org/https://doi.org/10.1016/S0375-9601(96)00706-2} {\bibfield
  {journal} {\bibinfo  {journal} {Physics Letters A}\ }\textbf {\bibinfo
  {volume} {223}},\ \bibinfo {pages} {1} (\bibinfo {year} {1996})}\BibitemShut
  {NoStop}%
\bibitem [{\citenamefont {Vidal}\ and\ \citenamefont
  {Werner}(2002)}]{PhysRevA.65.032314}%
  \BibitemOpen
  \bibfield  {author} {\bibinfo {author} {\bibfnamefont {G.}~\bibnamefont
  {Vidal}}\ and\ \bibinfo {author} {\bibfnamefont {R.~F.}\ \bibnamefont
  {Werner}},\ }\bibfield  {title} {\bibinfo {title} {Computable measure of
  entanglement},\ }\href {https://doi.org/10.1103/PhysRevA.65.032314}
  {\bibfield  {journal} {\bibinfo  {journal} {Phys. Rev. A}\ }\textbf {\bibinfo
  {volume} {65}},\ \bibinfo {pages} {032314} (\bibinfo {year}
  {2002})}\BibitemShut {NoStop}%
\bibitem [{\citenamefont {Plenio}(2005)}]{PhysRevLett.95.090503}%
  \BibitemOpen
  \bibfield  {author} {\bibinfo {author} {\bibfnamefont {M.~B.}\ \bibnamefont
  {Plenio}},\ }\bibfield  {title} {\bibinfo {title} {Logarithmic negativity: A
  full entanglement monotone that is not convex},\ }\href
  {https://doi.org/10.1103/PhysRevLett.95.090503} {\bibfield  {journal}
  {\bibinfo  {journal} {Phys. Rev. Lett.}\ }\textbf {\bibinfo {volume} {95}},\
  \bibinfo {pages} {090503} (\bibinfo {year} {2005})}\BibitemShut {NoStop}%
\bibitem [{\citenamefont {Bayat}\ \emph {et~al.}(2010)\citenamefont {Bayat},
  \citenamefont {Sodano},\ and\ \citenamefont {Bose}}]{PhysRevB.81.064429}%
  \BibitemOpen
  \bibfield  {author} {\bibinfo {author} {\bibfnamefont {A.}~\bibnamefont
  {Bayat}}, \bibinfo {author} {\bibfnamefont {P.}~\bibnamefont {Sodano}},\ and\
  \bibinfo {author} {\bibfnamefont {S.}~\bibnamefont {Bose}},\ }\bibfield
  {title} {\bibinfo {title} {Negativity as the entanglement measure to probe
  the kondo regime in the spin-chain kondo model},\ }\href
  {https://doi.org/10.1103/PhysRevB.81.064429} {\bibfield  {journal} {\bibinfo
  {journal} {Phys. Rev. B}\ }\textbf {\bibinfo {volume} {81}},\ \bibinfo
  {pages} {064429} (\bibinfo {year} {2010})}\BibitemShut {NoStop}%
\bibitem [{\citenamefont {Santos}\ \emph {et~al.}(2011)\citenamefont {Santos},
  \citenamefont {Korepin},\ and\ \citenamefont {Bose}}]{santos2011negativity}%
  \BibitemOpen
  \bibfield  {author} {\bibinfo {author} {\bibfnamefont {R.~A.}\ \bibnamefont
  {Santos}}, \bibinfo {author} {\bibfnamefont {V.}~\bibnamefont {Korepin}},\
  and\ \bibinfo {author} {\bibfnamefont {S.}~\bibnamefont {Bose}},\ }\bibfield
  {title} {\bibinfo {title} {Negativity for two blocks in the one-dimensional
  spin-1 affleck-kennedy-lieb-tasaki model},\ }\href
  {https://doi.org/10.1103/PhysRevA.84.062307} {\bibfield  {journal} {\bibinfo
  {journal} {Phys. Rev. A}\ }\textbf {\bibinfo {volume} {84}},\ \bibinfo
  {pages} {062307} (\bibinfo {year} {2011})}\BibitemShut {NoStop}%
\bibitem [{\citenamefont {Gray}\ \emph {et~al.}(2019)\citenamefont {Gray},
  \citenamefont {Bayat}, \citenamefont {Pal},\ and\ \citenamefont
  {Bose}}]{gray2019scale}%
  \BibitemOpen
  \bibfield  {author} {\bibinfo {author} {\bibfnamefont {J.}~\bibnamefont
  {Gray}}, \bibinfo {author} {\bibfnamefont {A.}~\bibnamefont {Bayat}},
  \bibinfo {author} {\bibfnamefont {A.}~\bibnamefont {Pal}},\ and\ \bibinfo
  {author} {\bibfnamefont {S.}~\bibnamefont {Bose}},\ }\bibfield  {title}
  {\bibinfo {title} {Scale invariant entanglement negativity at the many-body
  localization transition},\ }\href@noop {} {\bibfield  {journal} {\bibinfo
  {journal} {arXiv preprint arXiv:1908.02761}\ } (\bibinfo {year}
  {2019})}\BibitemShut {NoStop}%
\bibitem [{\citenamefont {Rajak}\ \emph {et~al.}(2023)\citenamefont {Rajak},
  \citenamefont {Suzuki}, \citenamefont {Dutta},\ and\ \citenamefont
  {Chakrabarti}}]{rajak2023quantum}%
  \BibitemOpen
  \bibfield  {author} {\bibinfo {author} {\bibfnamefont {A.}~\bibnamefont
  {Rajak}}, \bibinfo {author} {\bibfnamefont {S.}~\bibnamefont {Suzuki}},
  \bibinfo {author} {\bibfnamefont {A.}~\bibnamefont {Dutta}},\ and\ \bibinfo
  {author} {\bibfnamefont {B.~K.}\ \bibnamefont {Chakrabarti}},\ }\bibfield
  {title} {\bibinfo {title} {Quantum annealing: An overview},\ }\href@noop {}
  {\bibfield  {journal} {\bibinfo  {journal} {Philosophical Transactions of the
  Royal Society A}\ }\textbf {\bibinfo {volume} {381}},\ \bibinfo {pages}
  {20210417} (\bibinfo {year} {2023})}\BibitemShut {NoStop}%
\bibitem [{\citenamefont {Žnidarič}\ and\ \citenamefont
  {Ljubotina}(2018)}]{doi:10.1073/pnas.1800589115}%
  \BibitemOpen
  \bibfield  {author} {\bibinfo {author} {\bibfnamefont {M.}~\bibnamefont
  {Žnidarič}}\ and\ \bibinfo {author} {\bibfnamefont {M.}~\bibnamefont
  {Ljubotina}},\ }\bibfield  {title} {\bibinfo {title} {Interaction instability
  of localization in quasiperiodic systems},\ }\href
  {https://doi.org/10.1073/pnas.1800589115} {\bibfield  {journal} {\bibinfo
  {journal} {Proceedings of the National Academy of Sciences}\ }\textbf
  {\bibinfo {volume} {115}},\ \bibinfo {pages} {4595} (\bibinfo {year}
  {2018})},\ \Eprint
  {https://arxiv.org/abs/https://www.pnas.org/doi/pdf/10.1073/pnas.1800589115}
  {https://www.pnas.org/doi/pdf/10.1073/pnas.1800589115} \BibitemShut {NoStop}%
\bibitem [{\citenamefont {Lindblad}(1976)}]{Lindblad1976}%
  \BibitemOpen
  \bibfield  {author} {\bibinfo {author} {\bibfnamefont {G.}~\bibnamefont
  {Lindblad}},\ }\bibfield  {title} {\bibinfo {title} {On the generators of
  quantum dynamical semigroups},\ }\href {https://doi.org/10.1007/BF01608499}
  {\bibfield  {journal} {\bibinfo  {journal} {Communications in Mathematical
  Physics}\ }\textbf {\bibinfo {volume} {48}},\ \bibinfo {pages} {119}
  (\bibinfo {year} {1976})}\BibitemShut {NoStop}%
\bibitem [{\citenamefont {Medvedyeva}\ \emph {et~al.}(2016)\citenamefont
  {Medvedyeva}, \citenamefont {Prosen},\ and\ \citenamefont {\ifmmode
  \check{Z}\else \v{Z}\fi{}nidari\ifmmode~\check{c}\else
  \v{c}\fi{}}}]{PhysRevB.93.094205}%
  \BibitemOpen
  \bibfield  {author} {\bibinfo {author} {\bibfnamefont {M.~V.}\ \bibnamefont
  {Medvedyeva}}, \bibinfo {author} {\bibfnamefont {T.}~\bibnamefont {Prosen}},\
  and\ \bibinfo {author} {\bibfnamefont {M.}~\bibnamefont {\ifmmode
  \check{Z}\else \v{Z}\fi{}nidari\ifmmode~\check{c}\else \v{c}\fi{}}},\
  }\bibfield  {title} {\bibinfo {title} {Influence of dephasing on many-body
  localization},\ }\href {https://doi.org/10.1103/PhysRevB.93.094205}
  {\bibfield  {journal} {\bibinfo  {journal} {Phys. Rev. B}\ }\textbf {\bibinfo
  {volume} {93}},\ \bibinfo {pages} {094205} (\bibinfo {year}
  {2016})}\BibitemShut {NoStop}%
\bibitem [{\citenamefont {Fischer}\ \emph {et~al.}(2016)\citenamefont
  {Fischer}, \citenamefont {Maksymenko},\ and\ \citenamefont
  {Altman}}]{PhysRevLett.116.160401}%
  \BibitemOpen
  \bibfield  {author} {\bibinfo {author} {\bibfnamefont {M.~H.}\ \bibnamefont
  {Fischer}}, \bibinfo {author} {\bibfnamefont {M.}~\bibnamefont
  {Maksymenko}},\ and\ \bibinfo {author} {\bibfnamefont {E.}~\bibnamefont
  {Altman}},\ }\bibfield  {title} {\bibinfo {title} {Dynamics of a
  many-body-localized system coupled to a bath},\ }\href
  {https://doi.org/10.1103/PhysRevLett.116.160401} {\bibfield  {journal}
  {\bibinfo  {journal} {Phys. Rev. Lett.}\ }\textbf {\bibinfo {volume} {116}},\
  \bibinfo {pages} {160401} (\bibinfo {year} {2016})}\BibitemShut {NoStop}%
\bibitem [{\citenamefont {Rath}\ \emph {et~al.}(2023)\citenamefont {Rath},
  \citenamefont {Vitale}, \citenamefont {Murciano}, \citenamefont {Votto},
  \citenamefont {Dubail}, \citenamefont {Kueng}, \citenamefont {Branciard},
  \citenamefont {Calabrese},\ and\ \citenamefont
  {Vermersch}}]{PRXQuantum.4.010318}%
  \BibitemOpen
  \bibfield  {author} {\bibinfo {author} {\bibfnamefont {A.}~\bibnamefont
  {Rath}}, \bibinfo {author} {\bibfnamefont {V.}~\bibnamefont {Vitale}},
  \bibinfo {author} {\bibfnamefont {S.}~\bibnamefont {Murciano}}, \bibinfo
  {author} {\bibfnamefont {M.}~\bibnamefont {Votto}}, \bibinfo {author}
  {\bibfnamefont {J.}~\bibnamefont {Dubail}}, \bibinfo {author} {\bibfnamefont
  {R.}~\bibnamefont {Kueng}}, \bibinfo {author} {\bibfnamefont
  {C.}~\bibnamefont {Branciard}}, \bibinfo {author} {\bibfnamefont
  {P.}~\bibnamefont {Calabrese}},\ and\ \bibinfo {author} {\bibfnamefont
  {B.}~\bibnamefont {Vermersch}},\ }\bibfield  {title} {\bibinfo {title}
  {Entanglement barrier and its symmetry resolution: Theory and experimental
  observation},\ }\href {https://doi.org/10.1103/PRXQuantum.4.010318}
  {\bibfield  {journal} {\bibinfo  {journal} {PRX Quantum}\ }\textbf {\bibinfo
  {volume} {4}},\ \bibinfo {pages} {010318} (\bibinfo {year}
  {2023})}\BibitemShut {NoStop}%
\bibitem [{\citenamefont {Ghosh}\ and\ \citenamefont {\ifmmode \check{Z}\else
  \v{Z}\fi{}nidari\ifmmode~\check{c}\else
  \v{c}\fi{}}(2023)}]{ghoshPhysRevB.107.184303}%
  \BibitemOpen
  \bibfield  {author} {\bibinfo {author} {\bibfnamefont {R.}~\bibnamefont
  {Ghosh}}\ and\ \bibinfo {author} {\bibfnamefont {M.}~\bibnamefont {\ifmmode
  \check{Z}\else \v{Z}\fi{}nidari\ifmmode~\check{c}\else \v{c}\fi{}}},\
  }\bibfield  {title} {\bibinfo {title} {Relaxation of imbalance in a
  disordered xx model with on-site dephasing},\ }\href
  {https://doi.org/10.1103/PhysRevB.107.184303} {\bibfield  {journal} {\bibinfo
   {journal} {Phys. Rev. B}\ }\textbf {\bibinfo {volume} {107}},\ \bibinfo
  {pages} {184303} (\bibinfo {year} {2023})}\BibitemShut {NoStop}%
\bibitem [{\citenamefont {Beccaria}\ \emph {et~al.}(2006)\citenamefont
  {Beccaria}, \citenamefont {Campostrini},\ and\ \citenamefont
  {Feo}}]{PhysRevB.73.052402}%
  \BibitemOpen
  \bibfield  {author} {\bibinfo {author} {\bibfnamefont {M.}~\bibnamefont
  {Beccaria}}, \bibinfo {author} {\bibfnamefont {M.}~\bibnamefont
  {Campostrini}},\ and\ \bibinfo {author} {\bibfnamefont {A.}~\bibnamefont
  {Feo}},\ }\bibfield  {title} {\bibinfo {title} {Density-matrix
  renormalization-group study of the disorder line in the quantum axial
  next-nearest-neighbor ising model},\ }\href
  {https://doi.org/10.1103/PhysRevB.73.052402} {\bibfield  {journal} {\bibinfo
  {journal} {Phys. Rev. B}\ }\textbf {\bibinfo {volume} {73}},\ \bibinfo
  {pages} {052402} (\bibinfo {year} {2006})}\BibitemShut {NoStop}%
\bibitem [{\citenamefont {Suzuki}\ \emph {et~al.}(2013)\citenamefont {Suzuki},
  \citenamefont {Inoue},\ and\ \citenamefont {Chakrabarti}}]{Suzuki2013}%
  \BibitemOpen
  \bibfield  {author} {\bibinfo {author} {\bibfnamefont {S.}~\bibnamefont
  {Suzuki}}, \bibinfo {author} {\bibfnamefont {J.-i.}\ \bibnamefont {Inoue}},\
  and\ \bibinfo {author} {\bibfnamefont {B.~K.}\ \bibnamefont {Chakrabarti}},\
  }\bibinfo {title} {Annni model in transverse field},\ in\ \href
  {https://doi.org/10.1007/978-3-642-33039-1_4} {\emph {\bibinfo {booktitle}
  {Quantum Ising Phases and Transitions in Transverse Ising Models}}}\
  (\bibinfo  {publisher} {Springer Berlin Heidelberg},\ \bibinfo {address}
  {Berlin, Heidelberg},\ \bibinfo {year} {2013})\ pp.\ \bibinfo {pages}
  {73--103}\BibitemShut {NoStop}%
\bibitem [{\citenamefont {Haldar}\ \emph {et~al.}(2021)\citenamefont {Haldar},
  \citenamefont {Mallayya}, \citenamefont {Heyl}, \citenamefont {Pollmann},
  \citenamefont {Rigol},\ and\ \citenamefont {Das}}]{PhysRevX.11.031062}%
  \BibitemOpen
  \bibfield  {author} {\bibinfo {author} {\bibfnamefont {A.}~\bibnamefont
  {Haldar}}, \bibinfo {author} {\bibfnamefont {K.}~\bibnamefont {Mallayya}},
  \bibinfo {author} {\bibfnamefont {M.}~\bibnamefont {Heyl}}, \bibinfo {author}
  {\bibfnamefont {F.}~\bibnamefont {Pollmann}}, \bibinfo {author}
  {\bibfnamefont {M.}~\bibnamefont {Rigol}},\ and\ \bibinfo {author}
  {\bibfnamefont {A.}~\bibnamefont {Das}},\ }\bibfield  {title} {\bibinfo
  {title} {Signatures of quantum phase transitions after quenches in quantum
  chaotic one-dimensional systems},\ }\href
  {https://doi.org/10.1103/PhysRevX.11.031062} {\bibfield  {journal} {\bibinfo
  {journal} {Phys. Rev. X}\ }\textbf {\bibinfo {volume} {11}},\ \bibinfo
  {pages} {031062} (\bibinfo {year} {2021})}\BibitemShut {NoStop}%
\bibitem [{Note5()}]{Note5}%
  \BibitemOpen
  \bibinfo {note} {Since we work with just the connected correlation functions
  we sometimes may obtain large values of $\protect \mathcal {C}_1$ and
  $\protect \mathcal {C_2}$. Because we are mostly concerned with detection and
  not measurement we deem rescaling unnecessary. Bounds can however be found
  depending on the system size an operators involved. For the transverse Ising
  model in our work where we use total magnetization operator, the peak occurs
  at $h=0$ and takes a value $L^2/4$. This can be seen by computing the Gibbs
  state in the presence of infinitesimal longitudinal field at small but finite
  temperature--- $\protect \frac {1}{2}\mathinner {|{\uparrow \uparrow \protect
  \mathellipsis \uparrow }\rangle }\mathinner {\langle {\uparrow \uparrow
  \protect \mathellipsis \uparrow }|}+\protect \frac {1}{2}\mathinner
  {|{\downarrow \downarrow \protect \mathellipsis \downarrow }\rangle
  }\mathinner {\langle {\downarrow \downarrow \protect \mathellipsis \downarrow
  }|}$. This state gives the maximum possible unnormalized connected
  correlation for this observable, and in fact $\protect \mathcal {C}_2$ is
  loosely upper bounded by this value as well since we always rotate away from
  this special state.}\BibitemShut {Stop}%
\bibitem [{Note6()}]{Note6}%
  \BibitemOpen
  \bibinfo {note} {It is not separable to states with purely imaginary off
  diagonal elements either as taking $\protect \mathcal {O}=\DOTSB \sum@
  \slimits@ \langle |\sigma _z|\rangle $ does not yield $\protect \mathcal
  {C}_2=0$ in this region}\BibitemShut {NoStop}%
\bibitem [{\citenamefont {Meyer}\ and\ \citenamefont
  {Matveev}(2008)}]{Meyer_2009}%
  \BibitemOpen
  \bibfield  {author} {\bibinfo {author} {\bibfnamefont {J.~S.}\ \bibnamefont
  {Meyer}}\ and\ \bibinfo {author} {\bibfnamefont {K.~A.}\ \bibnamefont
  {Matveev}},\ }\bibfield  {title} {\bibinfo {title} {Wigner crystal physics in
  quantum wires},\ }\href {https://doi.org/10.1088/0953-8984/21/2/023203}
  {\bibfield  {journal} {\bibinfo  {journal} {Journal of Physics: Condensed
  Matter}\ }\textbf {\bibinfo {volume} {21}},\ \bibinfo {pages} {023203}
  (\bibinfo {year} {2008})}\BibitemShut {NoStop}%
\end{thebibliography}%

\appendix
\section{Properties of $U$ in Eq.~\eqref{eq:basis1}}
\label{app:appA}
We start with the following definition of $U$,
\begin{equation}
    U=(e^{-i \sigma^x \pi/4})^{\otimes L}
    \label{eq:basis11},
\end{equation}
where $L$ denotes the length of the subsystem and, 
\begin{equation*}
    \sigma_x=\begin{pmatrix}
        0 &1 \\
        1 &0
    \end{pmatrix}
\end{equation*}
For $L=1$, we have,
\begin{equation}
 U=   \left(
\begin{array}{cc}
 \frac{1}{\sqrt{2}} & -\frac{i}{\sqrt{2}} \\
 -\frac{i}{\sqrt{2}} & \frac{1}{\sqrt{2}} \\
\end{array}
\right)
\end{equation}
This can be written for future convenience as,
\begin{equation}
   U= \frac{1}{\sqrt{2}}\begin{pmatrix}
        1 & a \\ a & 1
    \end{pmatrix}
\label{eq:eqL1}
\end{equation}
with $a=e^{-i \pi/2}$.
Then for $L=2,3$ respectively we have,
\begin{eqnarray}
    U=\frac{1}{2}\left(
\begin{array}{cccc}
 1 & a & a & a^2 \\
 a & 1 & a^2 & a \\
 a & a^2 & 1 & a \\
 a^2 & a & a & 1 \\
\end{array}
\right), \nonumber \\
U=\frac{1}{2 \sqrt{2}}\left(
\begin{array}{cccccccc}
 1 & a & a & a^2 & a & a^2 & a^2 & a^3 \\
 a & 1 & a^2 & a & a^2 & a & a^3 & a^2 \\
 a & a^2 & 1 & a & a^2 & a^3 & a & a^2 \\
 a^2 & a & a & 1 & a^3 & a^2 & a^2 & a \\
 a & a^2 & a^2 & a^3 & 1 & a & a & a^2 \\
 a^2 & a & a^3 & a^2 & a & 1 & a^2 & a \\
 a^2 & a^3 & a & a^2 & a & a^2 & 1 & a \\
 a^3 & a^2 & a^2 & a & a^2 & a & a & 1 \\
\end{array}
\right)
\label{eq:genmat}
\end{eqnarray}
 Clearly $U=U^T$, which comes from $\sigma_x^T=\sigma_x$. Additionally we observe for these examples, $U_{p,q}U_{d-p+1,q}=\frac{1}{L}a^L$, where $d=2^L$. This can be verified for a general case as follows.

 Since we use the computational basis, the basis states in Eq.~\ref{eq:eqL1} will be $(\ket{0}, \ket{1})$, hence $U_{L=1}= \frac{1}{\sqrt{2}}(\ket{0}\bra{0}+\ket{1}\bra{1}+a\ket{0}\bra{1}+a\ket{1}\bra{0})$. Any element of $U$ for a generic $L$ can be constructed out of these elements due to the tensor product structure. For example in $L=3$, the $\ket{110}\bra{010}$ element can be constructed as $\bra{1}U_{L=1}\ket{0}\bra{1}U_{L=1}\ket{1}\bra{0}U_{L=1}\ket{0}=a\times1\times1=a$, which matches with the corresponding element in $U_{L=3}$ (row-$7$, column-$3$) in Eq.~\eqref{eq:genmat}. Now for an element of an arbitrary row index $p$, the basis state for the column indices $q$ and $d-q+1$ will be just bit flipped states of each other. For example, at $L=3$, if we consider $p=2$, then the $q=3$ and $q=6$ basis states would be $\ket{001}\bra{010}$ and $\ket{001}\bra{101}$. Then it can be seen that $U_{23} U_{26}=\frac{a^3}{2 \sqrt{2}}$ using Eq.~\eqref{eq:eqL1}. 
 
 It follows that,  $U_{p,q} U_{d-p+1,q}=\frac{1}{L}e^{-i \pi L/2}$, and any element of $U$ can be written as an integer power of $a$, i.e. 
 $U_{p,q}=\frac{1}{\sqrt{L}}e^{-i \pi(\phi_{p,q})/2}$, {\rm where} $\phi_{p,q} \in [0,L] \cap \mathbb{Z}$.
\section{An alternative unitary rotation}
\label{app:appB}
In this section we shall provide another example of the rotation which accomplishes the same task as the one described in the main text. While this rotation provides the advantage of applicability to qudit systems, it is significantly more non-local than the example in the main text, and thus harder to implement in practice.

The rotation matrix to be considered is closely related to the many-body Fourier transform local to the subsystems,
 \begin{equation}
 (U_A)_{pq}=\frac{1}{\sqrt{d}}(e^{\frac{2 i \pi (p-1/2) q}{d}})
 \label{eq:basis2}
 \end{equation}
and similarly for $U_B$. The purpose of adding the factor of $-2 i \pi q/2$ with the usual Fourier transform will be clear in the following computation. 
Using the specific nature of $\mathcal{O}$ we have,
\begin{eqnarray}
&&\Tr[\rho^{\prime} \mathcal{O}_A]=\Tr[\rho^{\prime A} \mathcal{O}_A] \nonumber \\
&&= \frac{1}{d}\sum_{p,q,r,s}^d e^{2 \pi i q \frac{p-1/2}{d}}\rho_{qr} e^{-2 \pi i r \frac{s-1/2}{d}} (f(s)+c_1) \delta_{sp} \nonumber \\
&=& \frac{1}{d}\sum_{p=1}^{d/2}\sum_{q,r}^d (f (p) +c_1) e^ {2 \pi i (\frac{p(q-r)}{d}-\frac{q-r}{2d})} \rho_{q r} \nonumber \\
&+& \frac{1}{d}\sum_{p=1}^{d/2}\sum_{q,r}^d (-[+]f (p) +c_1) e^ {2 \pi i (-\frac{p(q-r)}{d}+\frac{q-r}{2d})} \rho_{q r}
\label{eq:intermiediatefourier}
\end{eqnarray}
Then using the symmetry [antisymmetry] of $\rho$ (the first condition), i.e., $\rho_{qr}=[-]\rho_{rq}$ we can simplify Eq.~\ref{eq:intermiediatefourier} to obtain,
\begin{eqnarray}
\Tr[\rho^{\prime A} \mathcal{O}_A]&=&\frac{1}{d}\sum_{p=1}^{d}\sum_{q,r}^d c_1 e^ {2 \pi i (\frac{p(q-r)}{d}-\frac{q-r}{2d})} \rho_{q r} \nonumber \\
&&=c_1 \Tr[\rho^{\prime A}]=c_1
\label{eq:MUBvalue2}
\end{eqnarray}
It is worth noting here, that for odd $d$ Eq.~\eqref{eq:intermiediatefourier} will have an additional term and will be, 
\begin{eqnarray*}
&&\Tr[\rho^{\prime} \mathcal{O}_A]=\Tr[\rho^{\prime A} \mathcal{O}_A] \nonumber \\
&=& \frac{1}{d}\sum_{p=1}^{(d-1)/2}\sum_{q,r}^d (f (p) +c_1) e^ {2 \pi i (\frac{p(q-r)}{d}-\frac{q-r}{2d})} \rho_{q r} \nonumber \\
&+& \frac{1}{d}\sum_{p=1}^{(d-1)/2}\sum_{q,r}^d (-[+]f (p) +c_1) e^ {2 \pi i (-\frac{p(q-r)}{d}+\frac{q-r}{2d})} \rho_{q r}\nonumber \\
&+& \frac{1}{d}\sum_{q,r}^d (f ([d+1]/2) +c_1) e^ {2 \pi i (\frac{(d+1)(q-r)}{2d}-\frac{q-r}{2d})} \rho_{q r}
\end{eqnarray*}
Choosing $f([d+1]/2)=0$ gives us Eq.~\eqref{eq:MUBvalue2} back. This naturally arises for separable real matrices from $f(p)=-f(d-p+1)$, but becomes an additional condition for the purely imaginary off-diagonal elements case.
The rest of the proof follows as in the main text. Thus this rotation serves to obtain the appropriate observable in both even and odd dimensional Hilbert spaces.

\section{Proof of $\mathcal{P}_\mathcal{O}=1$ for states with conservation of total magnetization}
\label{app:appC}
Let us consider a mixed state of the form $\rho=\sum_i p_i |\psi_i\rangle\langle\psi_i|$. $|\psi_i\rangle$ can be any pure state with the constraint $\langle\psi_i|\sum_{j=1}^L\sigma^j_z|\psi_i\rangle=M$. 
We choose $\mathcal{O}_{A[B]}=\sum_{i=1}^{L_{A[B]}}\sigma^z_i$, hence the various terms of the Pearson correlation for this state $|\psi_i\rangle$ are,
\begin{eqnarray}
\langle\mathcal{O}_A\otimes \mathcal{O}_B\rangle&=&\sum_{k=0}^M|c_k|^2(M-k)k \label{a1} \\
\langle \mathcal{O}_A\otimes\mathbb{I}\rangle&=&(\sum_{k=0}^M(M-k)|c_k|^2)\label{a2}   \\
\langle \mathbb{I}\otimes\mathcal{O}_B\rangle&=&(\sum_{k=0}^Mk|c_k|^2)\label{a3}   \\
\langle\mathcal{O}_A^2\otimes \mathbb{I}\rangle&=&\sum_{k=0}^M(M-k)^2 |c_k|^2\label{a4}   \\
\langle\mathbb{I}\otimes \mathcal{O}_B^2\rangle&=&\sum_{k=0}^Mk^2 |c_k|^2 \label{a5} 
\end{eqnarray}
where $|c_k|^2$ is the probability of measuring $k$ spins up in subsystem $B$. 
Now $\mathcal{P}_{\mathcal{O}}$ can be written as,
\begin{widetext}
\begin{eqnarray}
  \mathcal{P}_{\mathcal{O}}&=&\frac{\langle\mathcal{O}_A\otimes \mathcal{O}_B\rangle-\langle \mathcal{O}_A\otimes\mathbb{I}\rangle\langle \mathbb{I}\otimes\mathcal{O}_B\rangle}{\sqrt{(\langle\mathcal{O}_A^2\otimes \mathbb{I}\rangle-\langle \mathcal{O}_A\otimes\mathbb{I}\rangle^2)(\langle\mathbb{I}\otimes \mathcal{O}_B^2\rangle-\langle \mathbb{I}\otimes\mathcal{O}_B\rangle^2)}} \nonumber \\
  &=&\frac{\sqrt{-[\sum_{k=0}^M(M-k) |c_k|^2]^2+M\sum_{k=0}^M(M-k)|c_k|^2-\sum_{k=0}^M(M-k)|c_k|^2\sum_{l=0}^M l |c_l|^2}}{\sqrt{\sum_{k=0}^M(M-k)^2 |c_k|^2-(\sum_{k=0}^M(M-k)|c_k|^2)^2}} \times \nonumber \\
 && \frac{\sqrt{-[\sum_{k=0}^M(M-k) |c_k|^2]^2+M\sum_{k=0}^Mk|c_k|^2-\sum_{l=0}^M(M-l)|c_l|^2\sum_{k=0}^M k |c_k|^2}}{\sqrt{\sum_{k=0}^M(M-k)^2 |c_k|^2-(\sum_{k=0}^M k|c_k|^2)^2}} =-1
  \end{eqnarray}
\end{widetext}

where we have used $\sum_{k=1}^M|c_k|^2=1$ and separately computed the ratios between the two terms in the denominator by introducing a square root in the numerator. This result is valid unless $(\langle\mathcal{O}_{A[B]}^2\otimes \mathbb{I}\rangle-\langle \mathbb{I}\otimes\mathcal{O}_{A[B]}\rangle^2$ yields $0$, i.e. the subsystem has a fixed number of particles as well.

By substituting this result for each $\ket{\psi_i}$ and using $Tr[\rho]=1$ we can show $P_{\mathcal{O}}=1$ for the full density matrix. Using the result of Ref.~\cite{Maccone1} it immediately follows that \textit{any correlation found in the MUB basis denotes the state is entangled}. Thus, measuring correlation in one basis is enough to detect entanglement.

\section{Full phase diagram of TI model}
\label{app:appD}

\begin{figure}
\centering
\includegraphics[width=0.85\columnwidth]{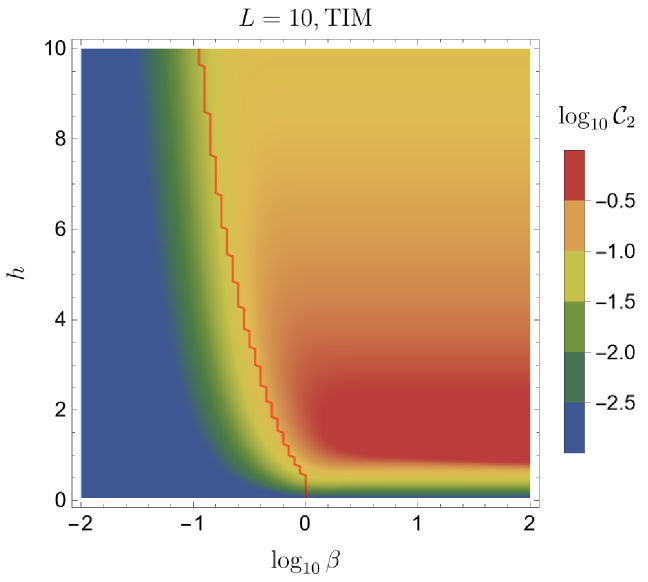}
\includegraphics[width=0.85\columnwidth]{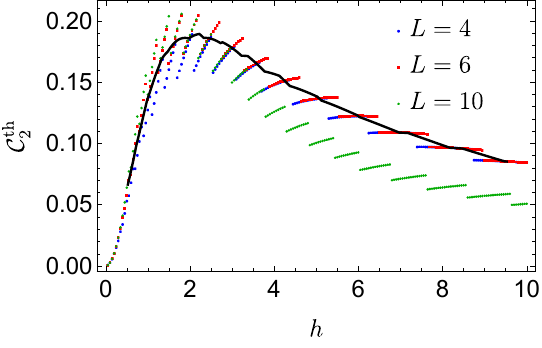}
\includegraphics[width=0.85\columnwidth]{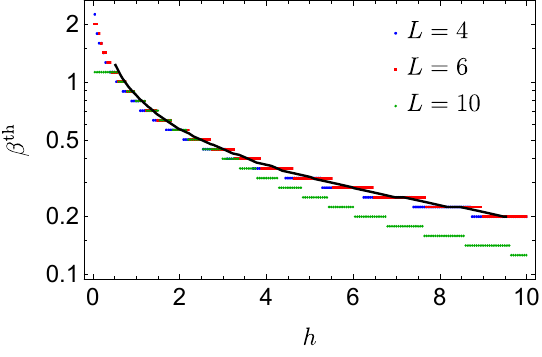}

\caption{Top:Behaviour of $\mathcal{C}_2$ for model in Eq.~\eqref{eq:ising} with $\kappa=0$ for different values if $\beta$ and $h$. The red line denotes the contour of $\mathcal{N}=10^{-4}$ to be used as an indicator to define lower bound of $\mathcal{C}_2$ to detect entanglement. Middle: Threshold $\mathcal{C}_2$ values which can be used to detect entanglement. Bottom:Threshold $\beta$ values above with $\mathcal{C}_2^{th}=0$ for different system sizes $L$. The black solid line denotes the moving average over $\delta h=0.1$}
\label{fig:fig3}
\end{figure}
In the top panel of Fig.~\ref{fig:fig3} we show the behaviour of $\mathcal{C}_2$ for a range of $\beta$ and $h$, while denoting the contour where $\mathcal{N}$ shows a non-zero value by red. Clearly for all values of $h$, $\mathcal{C}_2$ is an excellent witness of entanglement for $\beta>1$ i.e. at low temperature. For smaller $\beta$ we need to define a threshold value of $\mathcal{C}_2$, $\mathcal{C}_2^{th}$ to accurately detect entanglement. We show the numerically computed threshold values in the middle panel of Fig.~\ref{fig:fig3}. Two things are immediately apparent. The first is there is no significant finite size effect, in fact the threshold values reduce with system size which allows one to numerically compute the threshold for small system sizes and use them for experiments with a larger number of spins. Further accuracy can be achieved by a proper finite size analysis of the values which are beyond the scope of this work. The second prominent feature is the discrete `groupings' of the values, which occurs due to discrete grid of $h$ and the finite tolerance available to detect $\mathcal{N}$. Finally in the bottom panel of Fig.~\ref{fig:fig3}, we plot the threshold value of $\beta$, $\beta^{th}$ for different $h$, which is defined as the smallest value of $\beta$ which allows for accurate detection of entanglement with $\mathcal{C}_2^{th}=0$. This also shows reduction with system size $L$.
\begin{figure}
\centering
\includegraphics[width=0.45\columnwidth]{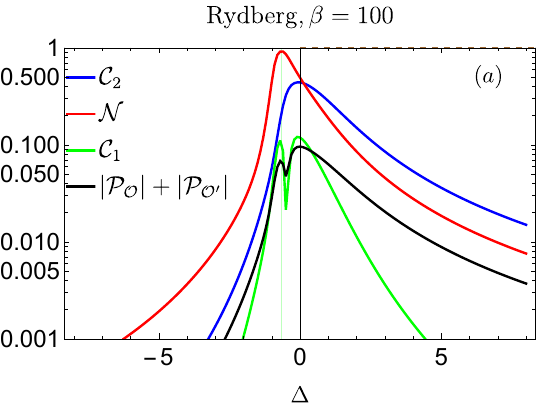}
\includegraphics[width=0.45\columnwidth]{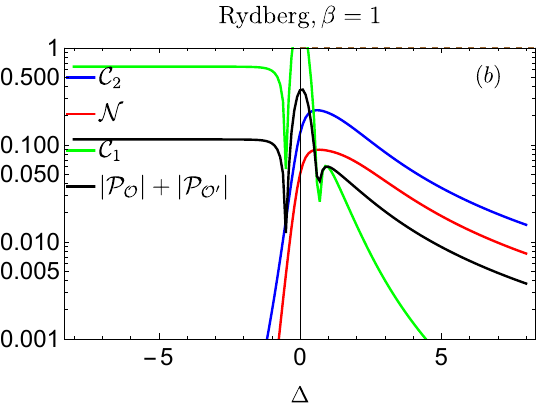}\\

\caption{Plots showing comparison of behaviour of $\mathcal{C}_2$ and $\mathcal{C}_1$ defined in Eqs.~\eqref{eq:C1} and \eqref{eq:C2}, to Negativity $\mathcal{N}$ and the criterion in Eq.~\eqref{eq:Macconecond} for the model given in Eq.~\eqref{eq:PXP}, (a) at $\beta=1/T=100$, (b) at $\beta=1$, The system size is $L=12$. The green grid line in panel (a) denotes the critical point.}
\label{fig:fig5}
\end{figure}
\section{Example III --- PXP model}
\label{app:appE}

As a final example,  we consider the kinetically constrained Rydberg atom model which has recently gained a lot of attention due to presence of `quantum scars'. In particular we study entanglement detection for states in thermal equilibrium [Eq.~\eqref{eq:thermal}] obtained from the Hamiltonian,
\begin{equation}
H=\sum_{i=1}^L (\Omega \tilde{\sigma}_i^x+\Delta \sigma_i^z) 
\label{eq:PXP}
\end{equation}
and show that our method has the potential to be applied for entanglement witnessing. Mapping an empty Rydberg site to a down spin and a filled site to an up spin, we denote here $\tilde{\sigma}_i^x = P\sigma_i^xP$ operation where $P$ denotes projection to only those states in Hilbert space with no neighbouring up spins. It is known that this model has a ferromagnetic ground state with all spins pointing down, i.e., presence of zero excited Rydberg atoms  when $\Delta \gg 0$ and $Z_2$ symmetry broken ground state for $\Delta \ll 0$.
As shown in Fig.~\ref{fig:fig5}, at large $\beta=100$, different correlators behave similarly.  In Fig.~\ref{fig:fig5}(a) we observe that while $\mathcal{N}$ shows a distinct peak at the known critical point of $\delta/\Omega=0.65$ and the maxima of $\mathcal{C}_2$ occurs at a slightly larger value. This is not unexpected as $\mathcal{C}_2$ is not an entanglement measure but a witness, hence we can only expect qualitative agreement. However, using Eq.~\eqref{eq:Macconecond} we cannot detect entanglement anywhere in the system as we never reach the threshold value of $1$. Expectedly, $\mathcal{C}_1$ shows completely different behaviour to $\mathcal{N}$ at low $\beta$ as seen Fig.~\ref{fig:fig5}(b) for a highly mixed state, where $\mathcal{C}_2$ qualitatively follows $\mathcal{N}$, but still there exists a region where $\mathcal{N}=0$ but $\mathcal{C}_2 \neq 0$. Note that the agreement between $\mathcal{C}_2$ and $\mathcal{N}$ is more pronounced in this case than the Ising model, thus allowing for smaller cut-offs and providing better detection at larger temperatures.
In Rydberg atoms, the change of basis can be simply executed experimentally via switching off the detuning field and the kinetic constraints by reducing the Rydberg blockade radius, and then taking a snapshot of the atomic density at different sites at $t=\frac{\pi}{4 h}+2 k \pi$,for $k \in \mathbb{Z}$, thus computing the correlation between the relevant subsystems.
\end{document}